%% file: main.tex
\documentclass[10pt,journal]{IEEEtran}
\IEEEoverridecommandlockouts
\usepackage{amsmath,amsfonts}
\usepackage{algorithmic}
\usepackage{algorithm}
\usepackage{array}
\usepackage[font=footnotesize]{caption} 
\usepackage{textcomp}
\usepackage{stfloats}
\usepackage{url}
\usepackage{verbatim}
\usepackage{subfigure}
\usepackage{graphicx,subcaption}
\usepackage{cite}
\usepackage{threeparttable}

\usepackage{xcolor}
\usepackage{multirow,booktabs}
\usepackage{makecell}
\usepackage{enumitem}
\usepackage{bm}
\usepackage{amssymb}
\usepackage{amsthm}

\usepackage{soul}

\theoremstyle{plain}

\allowdisplaybreaks[4]

\begin{document}

\bstctlcite{IEEEexample:BSTcontrol}

\title{Geometry-Aided Channel Deduction with Partial Channel Estimates and Uncalibrated Digital Twin}

\author{Hongning Ruan, Zhaoyang Zhang, Zirui Chen, Ziqing Xing, Zhaohui Yang, and Mérouane Debbah

\thanks{
Part of this work was presented at 2026 IEEE International Symposium on Personal, Indoor and Mobile Radio Communications (PIMRC) \cite{ruan2026gcd}.

This work was supported in part by National Natural Science Foundation of China under Grants 62394292 and 624B2129, and in part by the Fundamental Research Funds for the Central Universities under Grant 226-2024-00069. \textit{(Corresponding Author: Zhaoyang Zhang)}

H. Ruan, Z. Zhang, Z. Chen, Z. Xing, and Z. Yang are with the College of Information Science and Electronic Engineering, Zhejiang University, Hangzhou 310027, China, and also with Zhejiang Provincial Laboratory of Multi-Modal Communication Networks and Intelligent Information Processing, Hangzhou 310027, China. H. Ruan, Z. Zhang and Z. Xing are also with the Institute of Fundamental and Transdisciplinary Research, Zhejiang University, Hangzhou 310058, China. (e-mail: rhohenning@zju.edu.cn; ning\_ming@zju.edu.cn; ziruichen@zju.edu.cn; ziqing\_xing@zju.edu.cn; yang\_zhaohui@zju.edu.cn).

M. Debbah is with the Research Institute for Digital Future, Khalifa University, 127788 Abu Dhabi, UAE (email: merouane.debbah@ku.ac.ae).
}}

\maketitle

\begin{abstract}
The acquisition of high-dimensional channel state information (CSI) in wireless multi-input multi-output (MIMO) orthogonal frequency division multiplexing (OFDM) communications usually requires high pilot overhead, or relies on accurate and complete positional or environmental information. In this paper, we propose a geometry-aided channel deduction (GCD) approach, which utilizes an uncalibrated digital twin (DT) with only approximate environmental geometry and positions to assist the channel acquisition. The key rationale behind is that, even imprecise geometric information, which can be easily obtained in advance through radio sensing technologies or existing geographic databases, provides certain structural features about the current channel; meanwhile, the coarse instantaneous channel estimates using only a small amount of pilots provide dedicated information that aligns with the channel structure and further compensates for the geometry inaccuracy and other channel unknowns. To this end, we first extract geometric features from the DT, which contain only simple structural information of the channel. Then we propose \textit{random prompt augmentation}, a novel method to generate an appropriate prompt that converts geometric multi-path structure into a CSI-like representation while suppressing the disturbance of other unknown channel parameters. The prompt is then fused with the pilot-based instantaneous channel estimate via a channel deduction network. To further enhance the network's versatility, we incorporate pilot configurations into the existing learning architecture to support variable pilot patterns. Comprehensive experiments validate the superiority of the proposed method, which demonstrates high channel acquisition quality, low pilot overhead, and strong robustness. Furthermore, the structural prompt also serves as scenario-related context, enabling our approach to generalize well in new scenarios.
\end{abstract}

\begin{IEEEkeywords}
Channel acquisition, channel deduction, digital twin, ray tracing, cross-scenario generalization
\end{IEEEkeywords}

\section{Introduction}\label{sec::SecI}
\input{SecI}

\section{System Model}\label{sec::SecII}
\input{SecII}

\section{Proposed Framework}\label{sec::SecIII}
\input{SecIII}

\section{Simulation Results}\label{sec::SecIV}
\input{SecIV}

\section{Conclusion}\label{sec::SecV}
\input{SecV}


\bibliographystyle{IEEEtran}
\bibliography{ref.bib}

\end{document}

%% file: SecI.tex
\subsection{Background}

Future wireless networks are envisioned to evolve toward unprecedented levels of connectivity, data rate, spectral efficiency, and intelligence, enabled by a diverse range of emerging technologies, including extremely large-scale multiple-input multiple-output (MIMO) \cite{lu2014mimo}, terahertz communications, reconfigurable intelligent surfaces (RIS), stacked intelligent metasurfaces (SIM) \cite{zhang2026ssim1}, and integrated sensing and communication (ISAC) \cite{liu2022isac}. Along this evolution, the increasing scale of antenna arrays and wider transmission bandwidths are driving wireless channels toward higher dimensionality. Acquiring such high-dimensional channel state information (CSI) with low pilot overhead poses significant challenges.

Since the physical environments experienced by channels under different subcarriers and antennas are similar, a subset of the CSI can be mapped to the entire CSI, thereby reducing pilot overhead. However, the complexity of electromagnetic (EM) propagation characteristics makes it difficult for traditional interpolation methods to achieve optimal performance. Some methods \cite{schniter2014sparse,schniter2014gamp} estimate path parameters using pilot measurements to obtain complete CSI, but they typically involve high computational complexity for parameter estimation.
Deep learning excels at uncovering latent features and representing high-dimensional data, and has been applied to the field of channel estimation \cite{zhang2023aiextrapolation}. However, due to the dynamic and variable nature of wireless communication systems, such data-driven methods face challenges in terms of generalization. A key issue is whether deep neural networks can maintain their performance when channel propagation conditions change due to dynamic environmental variations or user mobility across different scenarios \cite{chen2025al}. Additionally, the neural network should be able to adapt to diverse system configurations and variable pilot patterns. These factors directly influence the feasibility of learning methods to be deployed in real-world systems.

\begin{table*}[t]
\setlength\tabcolsep{5pt}
\aboverulesep=0pt
\belowrulesep=0pt
\renewcommand{\arraystretch}{1.2}
\centering
\caption{Comparison of relevant works on channel acquisition.}
\label{tab::related_works}
\begin{tabular}{c|c|ccccc|c}
\toprule
Category & \makecell[c]{Auxiliary\\information} &
\makecell[c]{Low pilot\\overhead} &
\makecell[c]{Robustness to\\inaccurate\\information} &
\makecell[c]{No error\\propagation} &
\makecell[c]{Cross-scenario\\generalization\\capability} &
\makecell[c]{Low data\\collection cost} &
Related works \\
\midrule
\makecell[c]{Pilot-based channel estimation\\(pilot only)} & None &
& 
\checkmark &
\checkmark &
\checkmark &
\checkmark &
\cite{dong2019deepcnn,li2019reesnet,lin2021extrapolation,chen2024cmixer,zhou2025generative,arvinte2022score,fan2025ldm} \\
\midrule
\multirow{4}{*}{\makecell[c]{Pilot-free channel prediction\\(auxiliary information only)}} & Past channels &
\checkmark & 
&
&
\checkmark &
\checkmark &
\cite{jiang2022prediction,xiao2022odernn} \\
& Position &
\checkmark & 
&
\checkmark &
&
\checkmark &
\cite{xiao2022cgrbfnet,chatelier2025loc2ch} \\
& Real-time sensing &
\checkmark & 
&
\checkmark &
\checkmark &
&
\cite{liang2026multimodal} \\
& Radio environment &
\checkmark & 
&
\checkmark &
&
&
\cite{hoydis2024learning,jiang2025lwdt,an2025radiotwin,zhao2023nerf2,lu2024newrf,wen2025wrfgs} \\
\midrule
\multirow{5}{*}{\makecell[c]{Hybrid channel acquisition\\(pilot \& auxiliary information)}} & Past channels &
\checkmark & 
\checkmark &
&
\checkmark &
\checkmark &
\cite{chen2025cd} \\
& Position &
\checkmark & 
\checkmark &
\checkmark &
&
\checkmark &
\cite{chen2025scd} \\
& Real-time sensing &
\checkmark & 
\checkmark &
\checkmark &
\checkmark &
&
\cite{shi2025weicp} \\
& Channel dataset &
\checkmark & 
\checkmark &
\checkmark &
\checkmark &
&
\cite{chen2025scd,cai2026ecbp2wcp} \\
& Geometry &
\checkmark & 
\checkmark &
\checkmark &
\checkmark &
\checkmark &
Ours \\
\bottomrule
\end{tabular}
\end{table*}

\subsection{Related Works}

\subsubsection{Pilot-Based Channel Estimation}

Deep neural networks have been widely used for channel estimation. For example, \cite{dong2019deepcnn,li2019reesnet} employ convolutional neural network (CNN) to perform frequency domain interpolation, where pilot symbols are inserted into certain subcarriers. \cite{lin2021extrapolation} employs multi-layer perceptron (MLP) to perform spatial domain extrapolation, using CSI of a subset of antennas to infer that of other antennas. \cite{chen2024cmixer} proposed complex-domain MLP-Mixer (CMixer), which employs physics-inspired design to map partial estimates from a subset of antennas and subcarriers to complete spatial-frequency domain CSI, outperforming MLP- and CNN-based approaches. These neural networks are typically tailored to specific pilot configurations. Recent studies \cite{zhou2025generative,arvinte2022score,fan2025ldm} have explored the potential of diffusion models in channel estimation, using posterior sampling to support arbitrary pilot settings. However, such methods still require significant pilot resources to ensure channel estimation quality. In fact, when the multi-path characteristics of the channel are extremely complex, relying solely on limited pilot resources typically fails to provide sufficient channel features.

\subsubsection{Pilot-Free Channel Prediction}

Some studies have focused on achieving pilot-free channel prediction, based on the premise that some available information can be deterministically mapped to complete CSI. For example, the current channel can be predicted using channels from previous time slots \cite{jiang2022prediction,xiao2022odernn}. However, the small-scale mobility of users and subtle changes in the environment are often difficult to predict. Additionally, the auto-regressive inference accumulates prediction errors over time, leading to significant degradation in performance. Some studies \cite{xiao2022cgrbfnet,chatelier2025loc2ch} utilize user positions to predict channel information, which require extremely high positioning accuracy, as any positioning errors comparable to the wavelength can cause drastic changes in prediction results. Furthermore, since the wireless channel is jointly influenced by the user position and the surrounding environment, channel prediction models solely relying on positions lack generalization capabilities in new scenarios. \cite{liang2026multimodal} utilizes multi-modal sensing data to achieve adaptive CSI generation in dynamic environment, which ignores the impact of unknown environmental EM characteristics. Other studies focus on reconstructing the physical environment to create a digital twin (DT) that reflect the characteristics of wireless signal propagation. For example, \cite{hoydis2024learning,jiang2025lwdt,an2025radiotwin} utilize known environmental geometry and collect measured channel data to calibrate unknown environmental materials or learn EM interactions. The calibrated DT can be provided to a ray tracer to predict channel data at arbitrary positions. In contrast, \cite{zhao2023nerf2,lu2024newrf,wen2025wrfgs} focus on reconstructing a wireless radiation field based on measured channel data, without requiring environmental geometry. Despite their effectiveness, these methods still rely on high-precision user positioning, and the calibration or reconstruction process incurs data collection and computational costs for each individual scenario. Furthermore, all these methods overlook the diversity of antenna radiation characteristics, which is influenced by numerous factors such as hardware conditions and user device orientations.
As a result, when no pilot resources are available, it is practically only possible to obtain large-scale information -- such as received signal strength and channel gain \cite{sun2025channelgain1,sun2026channelgain2} -- rather than complete CSI. While such information is useful in certain applications, it cannot be used for more complex wireless tasks, such as symbol detection and precoding in multi-path scenarios.

\subsubsection{Hybrid Channel Acquisition}

Recent studies have enhanced channel estimation by combining pilot resources with auxiliary information. On one hand, the supplementary channel features provided by the auxiliary information significantly reduce pilot overhead; on the other hand, by introducing a small amount of pilot resources, these methods relax the accuracy requirements for the auxiliary information, as some small-scale or unpredictable features can be calibrated through instantaneous estimates from pilots. For example, \cite{chen2025cd} proposed the channel deduction (CD) framework that combines estimation with prediction, leveraging past channels to provide large-scale features to reduce pilot overhead. \cite{chen2025scd} proposed position-aided CMixer (PCMixer) that fuses position and channel information to obtain the complete channel, thereby enabling tolerance for positioning errors. To obtain more accurate and scenario-aware features, recent studies have incorporated scenario information to enhance performance and generalization capability. For example, \cite{shi2025weicp} extracts environmental features from multi-view images and combines it with partial CSI to obtain complete CSI. This method requires the mobile terminals to be equipped with sensing facilities and to continuously collect image data. \cite{chen2025scd} also proposed spatial channel deduction (SCD), which utilizes historical channel datasets from the current scenario to provide large-scale features. Similarly, \cite{cai2026ecbp2wcp} extracts channel knowledge from a full DT to enhance channel estimation. While they achieve outstanding performance, the acquisition of real-world channel data is expensive, whereas the full DT requires accurate and comprehensive environmental information, leading to high costs for deployment in new scenarios.

For clarity, we summarize relevant studies in Table \ref{tab::related_works}.

\subsection{Motivations and Contributions}

Recent advances in radio sensing technologies and existing geographic databases enable us to easily obtain static environmental maps, thereby supporting the construction of an approximate DT of the physical world. By utilizing the environmental geometry and transceiver positions, the multi-path structure of the channel can be obtained using a ray tracer \cite{alkhateeb2023dt,zhu2024dt}. Although these geometric details may contain inaccuracies, and many factors, such as environmental materials and antenna radiation characteristics, remain unobservable, the obtained structural features serve as effective prior information to enhance channel estimation, thereby significantly reducing pilot overhead. Meanwhile, geometric errors and other channel unknowns can be effectively compensated through instantaneous channel estimates using a small amount of pilots. Therefore, an uncalibrated DT without EM material properties is sufficient to effectively assist in channel acquisition. Furthermore, from the perspective of neural network training, geometric prior also introduces scenario-related context, enabling the neural network to extract common knowledge across different scenarios through multi-scenario collaborative learning, thereby demonstrating strong generalization capabilities in new scenarios.

Although our main idea shares the same spirit as recent studies that use DT \cite{del2025bayesian} or channel knowledge map (CKM) \cite{wang2025rekp} to enhance pilot-based channel estimation, it also differs significantly from these work in the following aspects. First, an ideal DT not only requires extremely high geometric accuracy with errors much smaller than the wavelength, but also entails substantial costs for data collection and computation to calibrate the unknown EM material properties \cite{hoydis2024learning}. In contrast, we use an approximate geometric map without specific material properties to provide only the channel structures and then generate appropriate prompts with them using a novel \textit{random prompt augmentation} mechanism, which suppresses the disturbance of other channel unknowns, such as the amplitude attenuation and phase shift of each path. Second, real-time DTs require online ray tracing to continuously update channel features, which incurs significant computational overhead and inference latency \cite{zhu2024dt}. In contrast, we employ only offline ray tracing to pre-extract channel features, and during online inference, we simply query them via neighborhood sampling based on user position, thereby avoiding the high costs associated with real-time ray tracing. Third, the CKM-based approaches typically maintain the CKM at the BS and require user positions for online CKM query \cite{zeng2024tutorial}, thus they are in general more suitable for uplink channel acquisition and may incur user privacy concerns. In contrast, our method stores and samples the pre-extracted channel features at the user end, which not only provides a solution for downlink CSI acquisition that usually involves more challenging spatial domain interpolation, but also completely circumvents the privacy issues.

The main contributions of this paper are as follows:

\begin{itemize}
\item We conduct a detailed theoretical analysis of the channel model, treat geometric path parameters as easily obtainable channel features, and propose the geometry-aided channel deduction (GCD) framework based on our analysis. This framework employs a channel deduction network to fuse geometric features extracted from the DT with pilot-based partial estimates, thereby deriving the complete channel.
\item We introduce random prompt augmentation, an innovative method to construct CSI prompts from simple structural information. This method not only converts geometric structural features into a complete CSI-like representation, helping the network fuse channel information from different modalities, but also suppresses other unknown features, enabling the network to fully leverage dominant structural information.
\item We present two neural network implementations for channel information fusion. One is GCDNet, which builds upon existing advanced learning architecture; the other is mGCDNet, which integrates pilot positional information to support variable pilot configurations. By enriching the training data diversity, mGCDNet improves versatility and enables more comprehensive learning.
\item We conduct extensive experiments to evaluate the proposed schemes. The results demonstrate their excellent channel acquisition accuracy and strong robustness against non-ideal geometric information. Furthermore, by incorporating multiple scenarios into the training process, we further enhance their overall learning performance and generalization capability.
\end{itemize}

The remainder of this paper is organized as follows. Section \ref{sec::SecII} introduces the channel model and analyzes the availability of various channel features. Based on this analysis, Section \ref{sec::SecIII} proposes the GCD framework and details its implementation. Section \ref{sec::SecIV} provides performance evaluation of proposed schemes. Finally, Section \ref{sec::SecV} concludes this paper.

%% file: SecII.tex
\subsection{Electromagnetic Wave Propagation}

This subsection formulates the channel response of a single-frequency wave propagating from a transmitting antenna to a receiving antenna \cite[Chapter 3]{asplund2020advanced}.

The direction-dependent radiation pattern of an antenna is defined as $\bm{g}(\widehat{\bm{k}})=g(\widehat{\bm{k}})\widehat{\bm{\psi}}(\widehat{\bm{k}})$, where $\widehat{\bm{k}}$ is a 3D unit vector representing the radiation direction, $g(\widehat{\bm{k}})$ is the complex amplitude gain, and $\widehat{\bm{\psi}}(\widehat{\bm{k}})$ represents the antenna polarization, i.e., the oscillation orientation of the transmitted field. $g(\widehat{\bm{k}})$ is normalized such that $g(\widehat{\bm{k}})=1$ corresponds to a lossless isotropic antenna\footnote{Formally, $g(\widehat{\bm{k}})$ must satisfy $\displaystyle\int_0^{2\pi}\int_0^{\pi}|g(\widehat{\bm{k}})|^2\sin\theta\mathrm{d}\theta\mathrm{d}\varphi=4\pi\eta$, where $\theta,\varphi$ are the zenith and azimuth angles defining the radiation direction $\widehat{\bm{k}}=[\sin\theta\cos\varphi,\sin\theta\sin\varphi,\cos\theta]^{\mathsf{T}}$, and $\eta$ is the antenna efficiency, i.e., the proportion of the input power that is converted into radiation.}. Under this definition, the electric far field of an antenna is proportional to $(1/d)e^{-jkd}\bm{g}(\widehat{\bm{k}})$, where $d$ is the propagation distance and $k$ is the wavenumber.

Generally, the propagation channel consists of $N_{\rm p}$ paths, where the $p$-th path has the following properties: path length $d_p$, departure direction $\widehat{\bm{k}}_{{\rm T},p}$ at the transmitter, arrival direction $\widehat{\bm{k}}_{{\rm R},p}$ at the receiver, and scattering matrix $\bm{\Xi}_p=\xi_p\bm{\Psi}_p\in\mathbb{C}^{3\times 3}$, where $\xi_p\in\mathbb{C}$ represents the attenuation and phase shift due to EM interactions between the wave and the scattering environment, and $\bm{\Psi}_p\in\mathbb{C}^{3\times 3}$ accounts for the relative attenuations and phase shifts between different polarization components of the field. For the line-of-sight (LoS) path, $\bm{\Xi}_p$ is the identity matrix, i.e., $\bm{\Xi}_p=\mathbf{I}$.

The channel frequency response can be expressed as
\begin{equation}\label{eq::h_simple}
h=\sum_{p=1}^{N_{\rm p}}\alpha_pe^{-j2\pi f\tau_p},
\end{equation}
where $f$ is the frequency, $\tau_p=d_p/c$ is the propagation delay of the $p$-th path, and $c$ denotes the speed of light. Let the radiation patterns of the transmitting and receiving antennas be $\bm{g}_{\rm T}(\widehat{\bm{k}})=g_{\rm T}(\widehat{\bm{k}})\widehat{\bm{\psi}}_{\rm T}(\widehat{\bm{k}})$ and $\bm{g}_{\rm R}(\widehat{\bm{k}})=g_{\rm R}(\widehat{\bm{k}})\widehat{\bm{\psi}}_{\rm R}(\widehat{\bm{k}})$, respectively. Then the path coefficient $\alpha_p\in\mathbb{C}$ is given by
\begin{equation}\label{eq::alpha_p}
\alpha_p=\frac{\lambda}{4\pi d_p}\bm{g}_{\rm R}(\widehat{\bm{k}}_{{\rm R},p})^{\mathsf{H}}\bm{\Xi}_p\bm{g}_{\rm T}(\widehat{\bm{k}}_{{\rm T},p}),
\end{equation}
which accounts for the propagation distance, antenna patterns, EM interactions between the wave and the environment, as well as the polarization mismatch between the incident wave and the receiving antenna.

\subsection{Channel Model}

We consider a MIMO system employing orthogonal frequency division multiplexing (OFDM), where a base station (BS) equipped with $N_{\rm t}$ antennas serves single-antenna users via $N_{\rm c}$ subcarriers. The downlink channel from the BS to a single user is represented by the spatial-frequency domain channel matrix $\mathbf{H}\in\mathbb{C}^{N_{\rm t}\times N_{\rm c}}$.

The frequency offset of the $n_{\rm c}$-th subcarrier is $\Delta f_{n_{\rm c}}=n_{\rm c}\Delta_{\rm f}$, where $\Delta_{\rm f}$ is the subcarrier spacing. Let $\bm{d}_{n_{\rm t}}$ denote the relative position of the $n_{\rm t}$-th BS antenna with respect to the local origin. Assuming that the wavefront at the BS is locally planar, the antenna's relative position results in the delay difference of $\Delta\tau_{n_{\rm t},p}=-\bm{d}_{n_{\rm t}}^{\mathsf{T}}\widehat{\bm{k}}_{{\rm T},p}/c$. By introducing frequency offsets and delay differences into \eqref{eq::h_simple}, the frequency response for the $n_{\rm t}$-th antenna and $n_{\rm c}$-th subcarrier is
\begin{align}
\mathbf{H}[n_{\rm t},n_{\rm c}]&=\sum_{p=1}^{N_{\rm p}}\alpha_pe^{-j2\pi(f+\Delta f_{n_{\rm c}})(\tau_p+\Delta\tau_{n_{\rm t},p})}\\
&\approx\sum_{p=1}^{N_{\rm p}}\alpha_pe^{-j2\pi f\tau_p}e^{-j2\pi n_{\rm c}\Delta_{\rm f}\tau_p}e^{jk\bm{d}_{n_{\rm t}}^{\mathsf{T}}\widehat{\bm{k}}_{{\rm T},p}}.\label{eq::H_element_sum}
\end{align}

To estimate the real-time channel, $N_{\rm t,\pi}$ antennas and $N_{\rm c,\pi}$ subcarriers are used to transmit non-precoded pilots, enabling the acquisition of the partial channel $\mathbf{H}_{\pi}=\mathbf{H}[\Omega]\in\mathbb{C}^{N_{\rm t,\pi}\times N_{\rm c,\pi}}$, where $\Omega=\Omega_{\rm t}\times\Omega_{\rm c}$ is a subset of antenna-subcarrier indices, and $\times$ denotes the Cartesian product. We assume that the pilots are evenly spaced in the $N_{\rm t}\times N_{\rm c}$ spatial-frequency grid. Therefore, the pilot pattern can be expressed in the following form:
\begin{subequations}
\begin{align}
\begin{aligned}
\Omega_{\rm t}=S_{\rm t}+\{0,R_{\rm t},\cdots,(N_{\rm t,\pi}-1)\times R_{\rm t}\},\\
S_{\rm t}\in\{1,2,\cdots,R_{\rm t}\},\quad R_{\rm t}=N_{\rm t}/N_{\rm t,\pi},
\end{aligned}\\
\begin{aligned}
\Omega_{\rm c}=S_{\rm c}+\{0,R_{\rm c},\cdots,(N_{\rm c,\pi}-1)\times R_{\rm c}\},\\
S_{\rm c}\in\{1,2,\cdots,R_{\rm c}\},\quad R_{\rm c}=N_{\rm c}/N_{\rm c,\pi},
\end{aligned}
\end{align}
\end{subequations}
where $R_{\rm t}$ and $S_{\rm t}$ denote the interval and the start of antenna indices in the spatial domain, while $R_{\rm c}$ and $S_{\rm c}$ denote those of subcarrier indices in the frequency domain.

Specifically, within each coherence time block, during which the channel state is approximately constant, the BS transmits pilot symbols on each subcarrier in $\Omega_{\rm c}$ through $N_{\rm q}$ transmissions ($N_{\rm q}\ge N_{\rm t,\pi}$). In the $q$-th transmission, the received signal at the user on the $n_{\rm c}$-th subcarrier ($n_{\rm c}\in\Omega_{\rm c}$) is given by $y_{q,n_{\rm c}}=\bm{h}_{n_{\rm c}}^{\mathsf{T}}\bm{x}_{q,n_{\rm c}}+w_{q,n_{\rm c}}$, where $\bm{h}_{n_{\rm c}}=\mathbf{H}[:,n_{\rm c}]\in\mathbb{C}^{N_{\rm t}}$ is the channel response on the $n_{\rm c}$-th subcarrier, $\bm{x}_{q,n_{\rm c}}\in\mathbb{C}^{N_{\rm t}}$ is the pilot vector for the $q$-th transmission, which satisfies $\|\bm{x}_{q,n_{\rm c}}\|_2^2=P$, with $P$ being the transmit power, and $w_{q,n_{\rm c}}\sim\mathcal{CN}(0,\sigma_{\rm w}^2)$ is the thermal noise, with $\sigma_{\rm w}^2$ being the thermal noise power. Based on a pilot transmission scheme similar to that in \cite{lee2021downlink}, we assume $N_{\rm q}=N_{\rm t,\pi}$, and on each subcarrier in $\Omega_{\rm c}$, the antennas in $\Omega_{\rm t}$ transmit pilot symbols successively through the $N_{\rm q}$ transmissions, with the $n_{\rm t}$-th antenna ($n_{\rm t}\in\Omega_{\rm t}$) corresponding to the $q(n_{\rm t})$-th transmission. In this case, each pilot vector $\bm{x}_{q,n_{\rm c}}$ has only one non-zero element $x_{q(n_{\rm t}),n_{\rm c}}$ corresponding to the $n_{\rm t}$-th antenna, and the channel $h_{n_{\rm t},n_{\rm c}}=\mathbf{H}[n_{\rm t},n_{\rm c}]$ at pilot positions can be simply estimated as $h_{n_{\rm t},n_{\rm c}}'=y_{q(n_{\rm t}),n_{\rm c}}/x_{q(n_{\rm t}),n_{\rm c}}=h_{n_{\rm t},n_{\rm c}}+\widetilde{w}_{q(n_{\rm t}),n_{\rm c}}$, where $\widetilde{w}_{q(n_{\rm t}),n_{\rm c}}=w_{q(n_{\rm t}),n_{\rm c}}/x_{q(n_{\rm t}),n_{\rm c}}\sim\mathcal{CN}(0,\widetilde{\sigma}_{\rm w}^2)$ is the estimation noise, $\widetilde{\sigma}_{\rm w}^2=\sigma_{\rm w}^2/P$ is the estimation noise power. In this paper, by default, we consider the ideal noise-free case where the clean partial CSI $\mathbf{H}_{\pi}$ can be obtained.

\subsection{Obtainable Channel Features}

We rewrite the MIMO-OFDM channel \eqref{eq::H_element_sum} as follows:
\begin{equation}\label{eq::H_element_dot}
\mathbf{H}[n_{\rm t},n_{\rm c}]=\bm{\phi}_{n_{\rm t},n_{\rm c}}^{\mathsf{T}}\widetilde{\bm{\alpha}},
\end{equation}
where $\bm{\phi}_{n_{\rm t},n_{\rm c}},\widetilde{\bm{\alpha}}\in\mathbb{C}^{N_{\rm p}}$ are $N_{\rm p}$-dimensional vectors, whose $p$-th elements are respectively defined as
\begin{align}
\phi_{n_{\rm t},n_{\rm c},p}&=e^{-j2\pi n_{\rm c}\Delta_{\rm f}\tau_p}e^{jk\bm{d}_{n_{\rm t}}^{\mathsf{T}}\widehat{\bm{k}}_{{\rm T},p}},\label{eq::phi_element_p}\\
\widetilde{\alpha}_p&=\alpha_pe^{-j2\pi f\tau_p}.\label{eq::tilde_alpha_p_simple}
\end{align}

The first component of the channel model, $\bm{\phi}_{n_{\rm t},n_{\rm c}}$, contains the phase offset caused by the $n_{\rm t}$-th antenna and $n_{\rm c}$-th subcarrier, which is determined by $\{(d_p,\widehat{\bm{k}}_{{\rm T},p})\}_{p=1}^{N_{\rm p}}$ and $(f,\Delta_{\rm f},\bm{d}_{n_{\rm t}})$. The path parameters $\{(d_p,\widehat{\bm{k}}_{{\rm T},p})\}_{p=1}^{N_{\rm p}}$, which serve as the geometric features describing the channel structure, can be obtained using ray tracing as long as the environmental map and the positions of the BS and the user are available. In practical applications, the BS can obtain the environmental map through radio sensing technologies or geographic information system (GIS), while the BS and the user can obtain their respective positions via global navigation satellite system (GNSS). However, for privacy reasons, the BS and users may be reluctant to disclose their positions externally. It should be noted that the acquired geometric information may exhibit errors that render the high-frequency phase term $e^{-j2\pi f\tau_p}$ unpredictable \cite{ruah2024calibrating}. Hence, we exclude $e^{-j2\pi f\tau_p}$ from the expression of $\bm{\phi}_{n_{\rm t},n_{\rm c}}$ and incorporate it into the second component $\widetilde{\bm{\alpha}}$. As for $(f,\Delta_{\rm f},\bm{d}_{n_{\rm t}})$, these are fixed system configurations determined by frequency band allocation and antenna array layout, and we assume they can be obtained with precision. Consequently, $\bm{\phi}_{n_{\rm t},n_{\rm c}}$ can be computed using the available information.

In contrast, the second component, $\widetilde{\bm{\alpha}}$, is typically unobtainable. To gain deeper insights into $\widetilde{\bm{\alpha}}$, we substitute $\alpha_p$ in \eqref{eq::tilde_alpha_p_simple} with \eqref{eq::alpha_p}, yielding the following complete expression:
\begin{equation}\label{eq::tilde_alpha_p_complete}
\widetilde{\alpha}_p=\frac{\lambda}{4\pi d_p}\bm{g}_{\rm R}(\widehat{\bm{k}}_{{\rm R},p})^{\mathsf{H}}\bm{\Xi}_p\bm{g}_{\rm T}(\widehat{\bm{k}}_{{\rm T},p})e^{-j2\pi f\tau_p},
\end{equation}
which is the product of several factors. The only factor that can be explicitly obtained is $\lambda/(4\pi d_p)$, which describes the large-scale attenuation determined by the propagation distance. The scattering matrix $\bm{\Xi}_p$ depends on the EM properties of environmental materials, while the antenna patterns $\bm{g}_{\rm T}(\widehat{\bm{k}}_{{\rm T},p})$ and $\bm{g}_{\rm R}(\widehat{\bm{k}}_{{\rm R},p})$ depend on hardware conditions and antenna orientations. These factors are typically difficult to obtain. Furthermore, as mentioned earlier, the high-frequency term $e^{-j2\pi f\tau_p}$ is unpredictable due to geometric errors. 

From the above analysis, it can be seen that $\bm{\phi}_{n_{\rm t},n_{\rm c}}$ represents the known structural features projected onto the $n_{\rm t}$-th antenna and $n_{\rm c}$-th subcarrier, whereas $\widetilde{\bm{\alpha}}$ represents unknown information shared across different antennas and subcarriers. Therefore, although the channel data has a high dimension of $O(N_{\rm t}N_{\rm c})$, the dimension of the underlying unknown features is quite low. Thus, by fully leveraging available geometric prior information, the unknown features can be compensated with minimal pilot overhead, thereby achieving complete channel acquisition.

%% file: SecIII.tex
\begin{figure}[t]
\vspace{-0.1cm}
\centering
\includegraphics[width=\columnwidth]{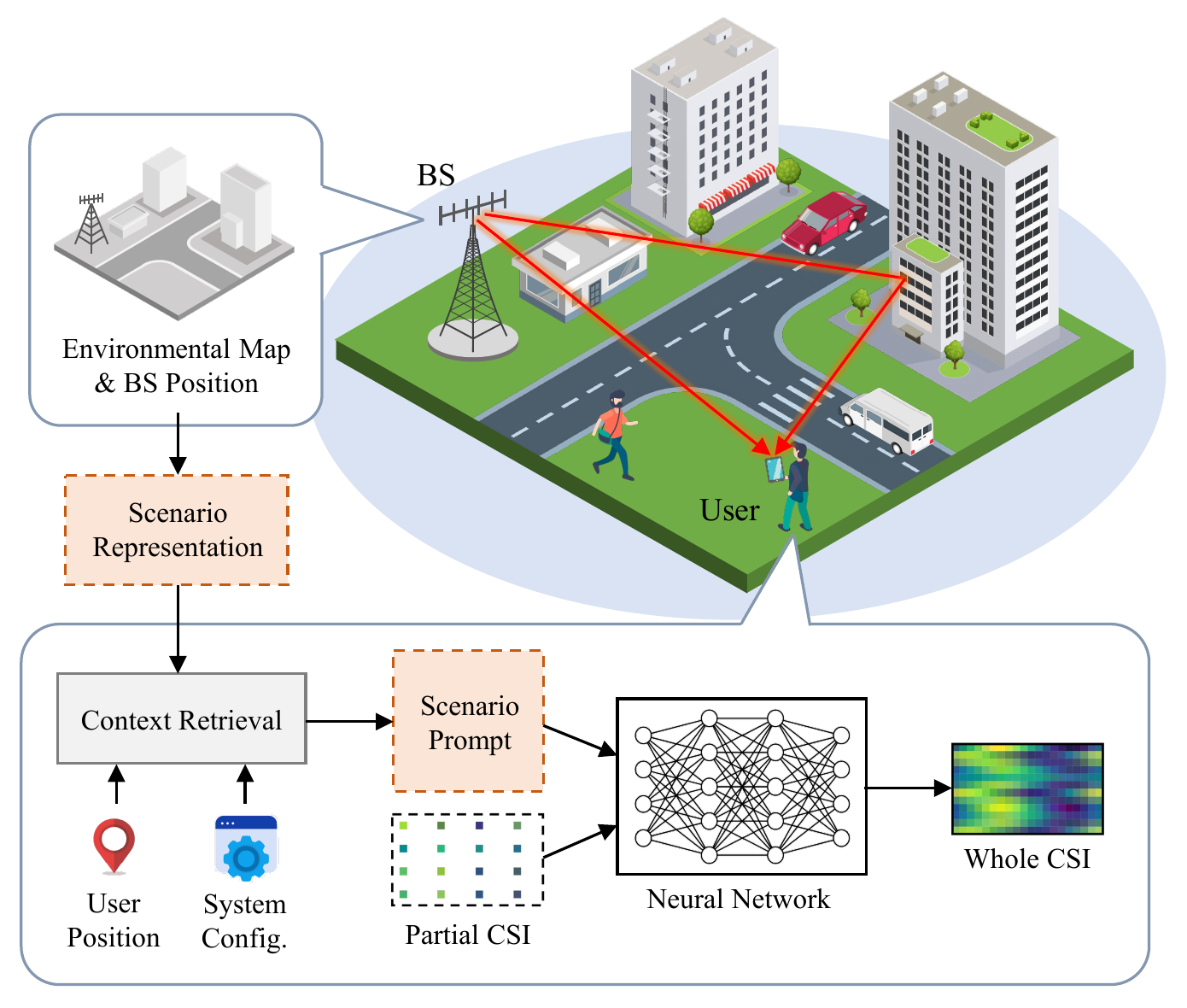}
\caption{Overview of the proposed channel acquisition framework.}
\label{fig::overview}
\vspace{-0.1cm}
\end{figure}

\begin{figure*}[t]
\vspace{-0.1cm}
\centering
\includegraphics[width=1.9\columnwidth]{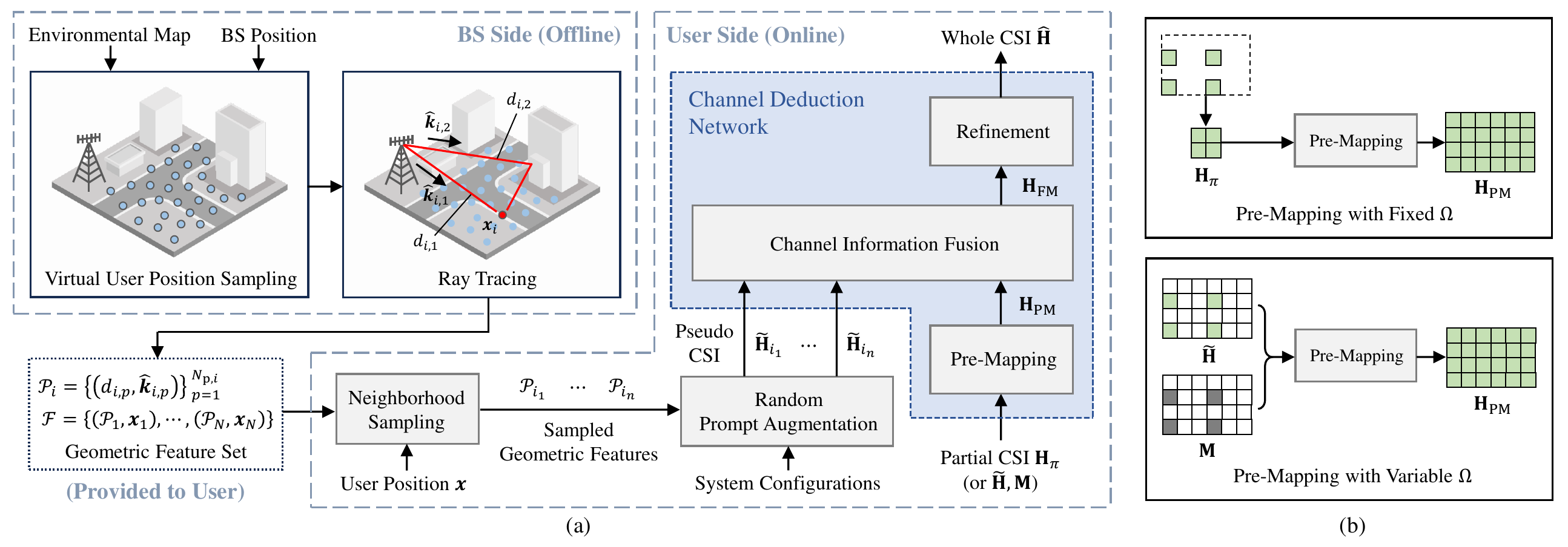}
\caption{(a) Detailed diagram of the proposed framework. (b) Comparison of the original pre-mapping module in \cite{chen2025cd} and the proposed pre-mapping module.}
\label{fig::method}
\vspace{-0.1cm}
\end{figure*}

To leverage available structural features and improve channel acquisition efficiency, we propose a novel framework named geometry-aided channel deduction (GCD). As shown in Fig. \ref{fig::overview}, the BS maintains an uncalibrated DT as the scenario representation, which consists of the BS position and an approximate environmental map. The user retrieves geometric features from the scenario representation based on its own position and converts them into the scenario prompt using known system configurations, thereby providing contextual information about the scenario. Meanwhile, the instantaneous channel is partially estimated by the user based on the received signal and known pilots. Finally, a neural network deployed at the user side fuses the partial channel estimate with the scenario prompt to generate the complete channel.

The following subsections detail the proposed framework, which is illustrated in Fig. \ref{fig::method}(a).

\subsection{Geometric Feature Extraction}

Although the multi-path structure can be obtained via ray tracing based on the environmental map and transceiver positions, there are several drawbacks to directly applying this method. First, ray tracing requires positional information from both the BS and the user, which raises privacy concerns because regardless of which one performs the ray tracing, the other must disclose its position. Furthermore, the user position is constantly changing, which means ray tracing must be repeatedly performed based on real-time user positions. This consumes substantial computational resources and results in significant inference latency \cite{zhu2024dt}.

Considering these issues, we enable the BS to pre-extract multi-path features offline without requiring actual user positions. We first sample $N$ virtual user positions $\bm{x}_1,\cdots,\bm{x}_N$ in the environmental map, ensuring these positions roughly cover the activity area of potential users. We then perform ray tracing using these positions to obtain the geometric feature set $\mathcal{F}=\{(\mathcal{P}_1,\bm{x}_1),\cdots,(\mathcal{P}_N,\bm{x}_N)\}$, where $\mathcal{P}_i=\{(d_{i,p},\widehat{\bm{k}}_{i,p})\}_{p=1}^{N_{{\rm p},i}}$ represents the channel structure at $\bm{x}_i$, $N_{{\rm p},i}$ denotes the number of paths from the BS to $\bm{x}_i$, and $d_{i,p},\widehat{\bm{k}}_{i,p}$ represent the length and departure direction of each path, respectively. Since $\mathcal{F}$ contains only path parameters at a finite number of discrete positions, $\mathcal{F}$ serves as a compact scenario representation with a small data volume and can be provisioned to the user during initial access and maintained at the user side. For multiple users within the same scenario, the same $\mathcal{F}$ can also be shared via broadcast or multicast, thereby reducing redundant transmission overhead.

Once the user has obtained the complete information of $\mathcal{F}$, the user is able to search for the nearest neighbors among the virtual user positions $\bm{x}_1,\cdots,\bm{x}_N$ within $\mathcal{F}$ based on its own position $\bm{x}$, with its positional information used only locally. We denote the $n$ nearest virtual user positions as $\bm{x}_{i_1},\cdots,\bm{x}_{i_n}$ and their corresponding channel structure as $\mathcal{P}_{i_1},\cdots,\mathcal{P}_{i_n}$. Given the high similarity of scattering environment within the spatial neighborhood, the structural features at position $\bm{x}$ can be well approximated by those at its neighbors.

\subsection{Random Prompt Augmentation}

Our proposed framework needs to fuse the geometric features and the partial CSI, which belong to different modalities and exhibit significant differences in data format. To ensure more learnable intra-modal fusion, we convert the geometric features $\mathcal{P}_i$ ($i=i_1,\cdots,i_n$) into the complete channel representation $\widetilde{\mathbf{H}}_i\in\mathbb{C}^{N_{\rm t}\times N_{\rm c}}$, which serves as the contextual prompt that can be directly fused with the partial CSI $\mathbf{H}_{\pi}$. This modality alignment process appears similar to the process in vision-language models that converts images into the text representation (i.e., tokens). However, unlike the semantic relationship between text and images that is primarily determined by human cognition, the relationship between path parameters and channel responses is strictly constrained by the physical laws \cite{chen2025towards}. Simply applying a learnable module to transform these two modalities into the same feature space cannot guarantee compliance with the physical constraints, thereby reducing learning efficiency. Instead, we propose random prompt augmentation, which manually computes a pseudo channel $\widetilde{\mathbf{H}}_i$ based on the channel model using the path parameters $\mathcal{P}_i$, thus strictly adhering to the physical constraints and perfectly transforming the path parameters into a channel representation. The subsequent neural network can process the pseudo channel the same as real channels, thereby achieving seamless modality alignment.

We assume that $\widetilde{\mathbf{H}}_i[n_{\rm t},n_{\rm c}]$ has a form similar to that in \eqref{eq::H_element_dot}, i.e., $\bm{\phi}_{i,n_{\rm t},n_{\rm c}}^{\mathsf{T}}\widetilde{\bm{\alpha}}_i$. Here, the first component $\bm{\phi}_{i,n_{\rm t},n_{\rm c}}\in\mathbb{C}^{N_{{\rm p},i}}$ is obtainable given the multi-path structure and system configurations, and its $p$-th element can be calculated as in \eqref{eq::phi_element_p}:
\begin{equation}\label{eq::phi_i_element_p}
\phi_{i,n_{\rm t},n_{\rm c},p}=e^{-j2\pi n_{\rm c}\Delta_{\rm f}\tau_{i,p}}e^{jk\bm{d}_{n_{\rm t}}^{\mathsf{T}}\widehat{\bm{k}}_{i,p}},
\end{equation}
and $\tau_{i,p}=d_{i,p}/c$. Based on our previous analysis of \eqref{eq::tilde_alpha_p_complete}, each element $\widetilde{\alpha}_{i,p}$ of the second component $\widetilde{\bm{\alpha}}_i$ is the product of a large-scale attenuation factor $\lambda/(4\pi d_{i,p})$ and a series of unknown factors. To handle the unknown channel parameters in $\widetilde{\bm{\alpha}}_i$, we introduce a placeholder vector $\widetilde{\bm{z}}_i\in\mathbb{C}^{N_{{\rm p},i}}$ to replace $\widetilde{\bm{\alpha}}_i$, and compute $\widetilde{\mathbf{H}}_i$ as follows:
\begin{equation}
\widetilde{\mathbf{H}}_i[n_{\rm t},n_{\rm c}]=\bm{\phi}_{i,n_{\rm t},n_{\rm c}}^{\mathsf{T}}\widetilde{\bm{z}}_i,
\end{equation}
where each element of $\widetilde{\bm{z}}_i$ is modeled as a random variable obeying the complex Gaussian distribution:
\begin{equation}\label{eq::tilde_z_element}
\widetilde{z}_{i,p}=\frac{\lambda}{4\pi d_{i,p}}z_{i,p},\quad z_{i,p}\sim\mathcal{CN}(0,\sigma_{\rm z}^2),
\end{equation}
and $\sigma_{\rm z}$ is a hyperparameter that controls the magnitude scale of the unknown factors. Our method is robust to the selection of $\sigma_{\rm z}$. In our experiments, $\sigma_{\rm z}$ is set to 0.5.

The random placeholders introduced here serve multiple purposes. First, they supplement unknown information within the channels, converting known multi-path structure into the complete channel representation. This enables the subsequent neural network to fuse different channel information within the same feature space. Second, by introducing different realizations of the random placeholder for each pseudo channel, the neural network suppresses the influence of these unknown channel parameters, thereby focusing on exploiting useful known structural information shared by all pseudo channels. This is because the neural network is unlikely to learn regular patterns from random data, thus avoiding the learning of spurious correlations between the placeholders and the real channel. This strategy is analogous to data augmentation in neural network training, which prevents the neural network from overfitting to unimportant features within data samples.
However, in our method, data augmentation is not applied to training samples but rather to the input prompt, enabling the network to identify important features within the contextual information.

\begin{figure*}[t]
\vspace{-0.1cm}
\centering
\includegraphics[width=1.9\columnwidth]{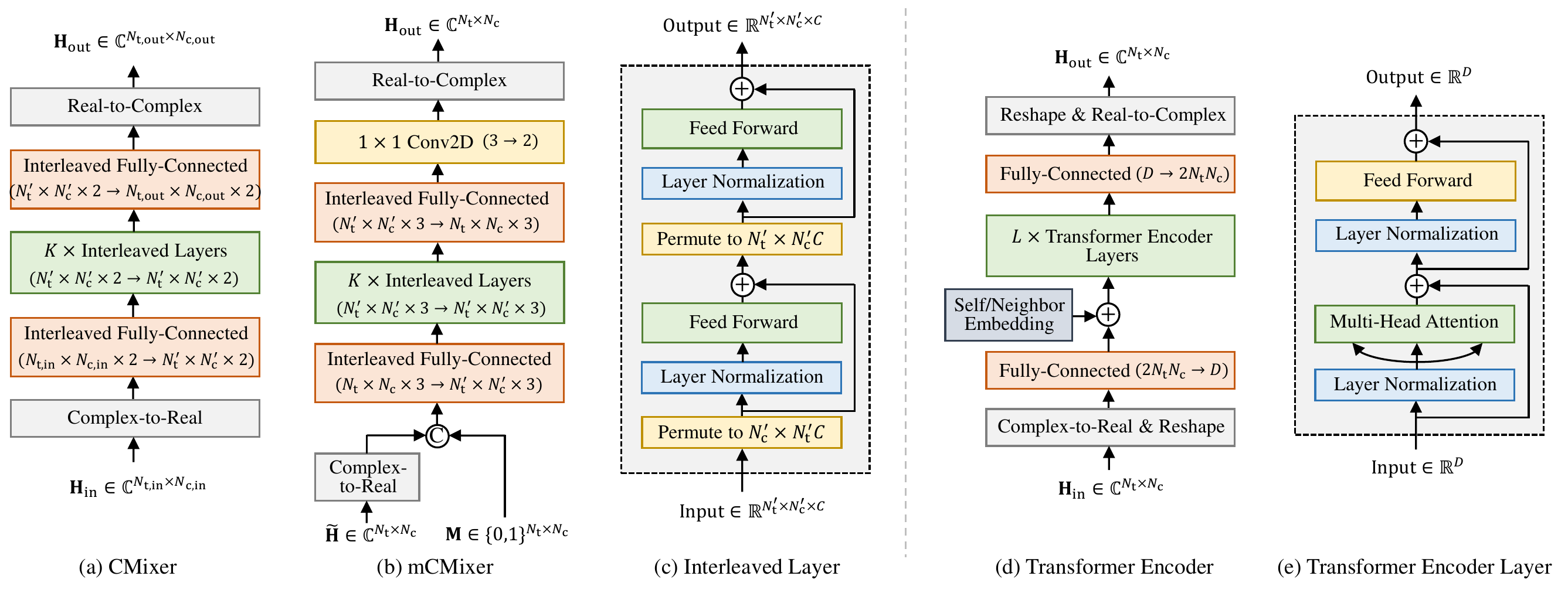}
\caption{Specific structures of the modules used in the channel deduction network. The plus sign ``+" denotes residual connection, and ``C" denotes concatenation.}
\label{fig::network}
\vspace{-0.1cm}
\end{figure*}

\subsection{Channel Deduction Network}

\subsubsection{GCDNet}

Following \cite{chen2025cd}, we fuse the pseudo channels with the partial channel using a channel deduction network and refer to it as GCDNet. It first pre-maps the partial CSI $\mathbf{H}_{\pi}$ to the form of a complete CSI $\mathbf{H}_{\rm PM}\in\mathbb{C}^{N_{\rm t}\times N_{\rm c}}$. Then, the network fuses features from $\widetilde{\mathbf{H}}_{i_1},\cdots,\widetilde{\mathbf{H}}_{i_n}$ and $\mathbf{H}_{\rm PM}$ to obtain a more informative representation $\mathbf{H}_{\rm FM}\in\mathbb{C}^{N_{\rm t}\times N_{\rm c}}$. Finally, $\mathbf{H}_{\rm FM}$ is further refined to yield the final output $\widehat{\mathbf{H}}$.

The channel pre-mapping module and the refinement module process channel data in spatial-frequency domain. Both modules are implemented using CMixer \cite{chen2024cmixer}, whose network structure is illustrated in Fig. \ref{fig::network}(a). CMixer treats the input $\mathbf{H}_{\rm in}\in\mathbb{C}^{N_{\rm t,in}\times N_{\rm c,in}}$ and the output $\mathbf{H}_{\rm out}\in\mathbb{C}^{N_{\rm t,out}\times N_{\rm c,out}}$ as real-valued tensors of shapes $N_{\rm t,in}\times N_{\rm c,in}\times 2$ and $N_{\rm t,out}\times N_{\rm c,out}\times 2$, respectively. It comprises $K$ stacked interleaved layers for channel representation learning, as well as two interleaved linear projections for dimension transformation. The detailed structure of the interleaved layer is shown in Fig. \ref{fig::network}(c). Its unique interleaved learning design closely aligns with the intrinsic structures of channel data, enabling CMixer to outperform existing learning methods in channel-related tasks with high parameter efficiency. In the channel deduction network, a $K$-layer CMixer is used to implement the pre-mapping module $\mathrm{PM}:\mathbb{C}^{N_{\rm t,\pi}\times N_{\rm c,\pi}}\to\mathbb{C}^{N_{\rm t}\times N_{\rm c}}$, converting the partial CSI $\mathbf{H}_{\pi}$ to $\mathbf{H}_{\rm PM}=\mathrm{PM}(\mathbf{H}_{\pi})$. Meanwhile, another $K$-layer CMixer serves as the refinement module $\mathrm{RM}:\mathbb{C}^{N_{\rm t}\times N_{\rm c}}\to\mathbb{C}^{N_{\rm t}\times N_{\rm c}}$, generating the output $\widehat{\mathbf{H}}=\mathrm{RM}(\mathbf{H}_{\rm FM})$.

The channel information fusion module is implemented using attention mechanism, which enables mutual interaction of channel features from $\widetilde{\mathbf{H}}_{i_1},\cdots,\widetilde{\mathbf{H}}_{i_n}$ and $\mathbf{H}_{\rm PM}$. The detailed network structure is illustrated in Fig. \ref{fig::network}(d)-(e). The input channels are first reshaped and compressed via linear projection, transforming them into $D$-dimensional real-valued vectors. Learnable embeddings are added to these vectors to distinguish the features from instantaneous estimate or pseudo channels. Subsequently, these vectors are fed into an $L$-layer Transformer encoder for information interaction. Finally, the last output vector corresponding to $\mathbf{H}_{\rm PM}$ is transformed to the output channel via linear projection and reshaping. The entire process is expressed as $\mathbf{H}_{\rm FM}=\mathrm{FM}(\widetilde{\mathbf{H}}_{i_1},\cdots,\widetilde{\mathbf{H}}_{i_n},\mathbf{H}_{\rm PM})$, where $\mathrm{FM}:\mathbb{C}^{N_{\rm t}\times N_{\rm c}}\times\cdots\times\mathbb{C}^{N_{\rm t}\times N_{\rm c}}\to\mathbb{C}^{N_{\rm t}\times N_{\rm c}}$ denotes the fusion module.

\subsubsection{Mask-Enhanced GCDNet}

The above implementation of the pre-mapping module takes only partial CSI $\mathbf{H}_{\pi}$ as input, ignoring information about the pilot settings. Consequently, the entire network can only be applied to a fixed pilot pattern $\Omega$, i.e., fixed intervals $R_{\rm t},R_{\rm c}$ and fixed starts $S_{\rm t},S_{\rm c}$.

To enhance the network's universality, we incorporate pilot positional information into the pre-mapping module. We define the pilot mask $\mathbf{M}\in\{0,1\}^{N_{\rm t}\times N_{\rm c}}$ as follows, which indicates the pilot positions in the spatial-frequency grid.
\begin{equation}
\mathbf{M}[n_{\rm t},n_{\rm c}]=\begin{cases}
1,&(n_{\rm t},n_{\rm c})\in\Omega,\\
0,&(n_{\rm t},n_{\rm c})\not\in\Omega.
\end{cases}
\end{equation}
According to the pilot positions, the partial CSI $\mathbf{H}_{\pi}$ can be padded with zeros to obtain a sparse channel matrix $\widetilde{\mathbf{H}}=\mathbf{H}\odot\mathbf{M}\in\mathbb{C}^{N_{\rm t}\times N_{\rm c}}$, where $\odot$ denotes the Hadamard product.

We extend the network structure for pre-mapping, as shown in Fig. \ref{fig::network}(b). The padded CSI $\widetilde{\mathbf{H}}$ is first converted to a real-valued tensor of shape $N_{\rm t}\times N_{\rm c}\times 2$, and then concatenated with the mask $\mathbf{M}$ to obtain a composite tensor of shape $N_{\rm t}\times N_{\rm c}\times 3$. Then it passes through $K$ interleaved learning layers, which are quite similar to the original CMixer. Finally, we employ $1\times 1$ convolution with 3 input channels and 2 output channels to transform the data shape from $N_{\rm t}\times N_{\rm c}\times 3$ into $N_{\rm t}\times N_{\rm c}\times 2$. We refer to this extended structure as mask-enhanced CMixer (mCMixer). The output is expressed as $\mathbf{H}_{\rm PM}=\mathrm{mPM}(\widetilde{\mathbf{H}},\mathbf{M})$, where $\mathrm{mPM}:\mathbb{C}^{N_{\rm t}\times N_{\rm c}}\times\{0,1\}^{N_{\rm t}\times N_{\rm c}}\to\mathbb{C}^{N_{\rm t}\times N_{\rm c}}$ denotes the extended pre-mapping module.

By replacing the original pre-mapping module with a $K$-layer mCMixer, the extended GCDNet is capable of processing partial channels with various pilot patterns. A comparison between the original pre-mapping module and the extended pre-mapping module is illustrated in Fig. \ref{fig::method}(b). We refer to this extended network as mask-enhanced GCDNet (mGCDNet).

\subsection{Training and Inference}

The training of (m)GCDNet requires a dataset $\mathcal{D}$ collected in the communication scenario, where each data sample $(\mathbf{H},\bm{x})$ contains the complete CSI and user position. Additionally, this scenario should provide a geometric feature set $\mathcal{F}$ as the scenario representation. Thanks to the scenario-related context introduced by the geometric prior, (m)GCDNet supports strong adaptation and generalization across various scenarios. Therefore, we adopt multi-scenario collaborative training to leverage the rich data from different scenarios for more comprehensive and thorough learning. Assume $M$ different scenarios are involved in training, each with its own training dataset and geometric feature set. For the special case when $M=1$, collaborative learning reduces to single-scenario learning, a quite common setting in existing studies. Furthermore, after introducing the mask-enhanced pre-mapping module, mGCDNet is capable of handling different pilot patterns. We introduce multiple pilot configurations $\{\Omega_{\ell}\}_{\ell}$ during training to enhance the network's universality and enable more thorough learning.

At each training step, we uniformly sample a scenario index $m$ from $\{1,\cdots,M\}$, and then sample a CSI-position pair $(\mathbf{H},\bm{x})$ from the $m$-th scenario's dataset. Meanwhile, we sample $\Omega$ from all pre-defined pilot patterns $\{\Omega_{\ell}\}_{\ell}$. Using the complete CSI $\mathbf{H}$ and the pilot pattern $\Omega$, we can derive the partial CSI $\mathbf{H}_{\pi}$, the padded CSI $\widetilde{\mathbf{H}}$, and the pilot mask $\mathbf{M}$, which serve as inputs to the channel deduction network. Subsequently, we follow the aforementioned procedure to acquire the complete CSI $\widehat{\mathbf{H}}$. To mitigate the impact of the absolute magnitude of channel responses, we normalize the CSI data before feeding it into the network. Specifically, we first use known partial CSI to estimate the average power, denoted as $P_{\mathbf{H}}=\|\widetilde{\mathbf{H}}\|_{\rm F}^2/\|\mathbf{M}\|_1$, and then normalize each input channel matrix $\mathbf{H}$ via $\mathbf{H}\gets\mathbf{H}/\sqrt{P_{\mathbf{H}}}$. This ensures that the network inputs have a unit average power, which helps stabilize the training process and reduces the learning burden. The network is trained using the Adam optimizer, and the loss function is defined as the mean squared error (MSE) between the normalized $\mathbf{H},\widehat{\mathbf{H}}$, i.e., $\mathcal{L}=\|\mathbf{H}-\widehat{\mathbf{H}}\|_{\rm F}^2/(N_{\rm t}N_{\rm c})$.

During inference, we normalize the network inputs using the estimated magnitude $\sqrt{P_{\mathbf{H}}}$, similar to the training phase. To recover the acquired channel with absolute magnitude scale, the output CSI can be denormalized via $\widehat{\mathbf{H}}\gets\sqrt{P_{\mathbf{H}}}\widehat{\mathbf{H}}$.

Since ray tracing and feature extraction are performed offline in our approach, online inference only requires neighborhood search, pseudo CSI construction, and the neural network forward pass. Assuming that neighborhood search is implemented by identifying $n$ minimum elements among the $N$ distances between $\bm{x}$ and each of $\bm{x}_1,\cdots,\bm{x}_N$, the overall computational complexity for neighborhood search is $O(N+N\log n)=O(N\log n)$. The complexity for pseudo CSI construction is $O(N_{\rm t}N_{\rm c}N_{\rm p}n)$, where $N_{\rm p}$ is the maximum number of paths, while the forward pass of (m)GCDNet has a computational complexity of $O(K(N_{\rm c}N_{\rm t}^2+N_{\rm t}N_{\rm c}^2)+N_{\rm t}N_{\rm c}Dn+L(D^2n+Dn^2))$ \cite{chen2025cd}.

%% file: SecIV.tex
\begin{figure}[t]
\vspace{-0.1cm}
\centering
\includegraphics[width=0.9\columnwidth]{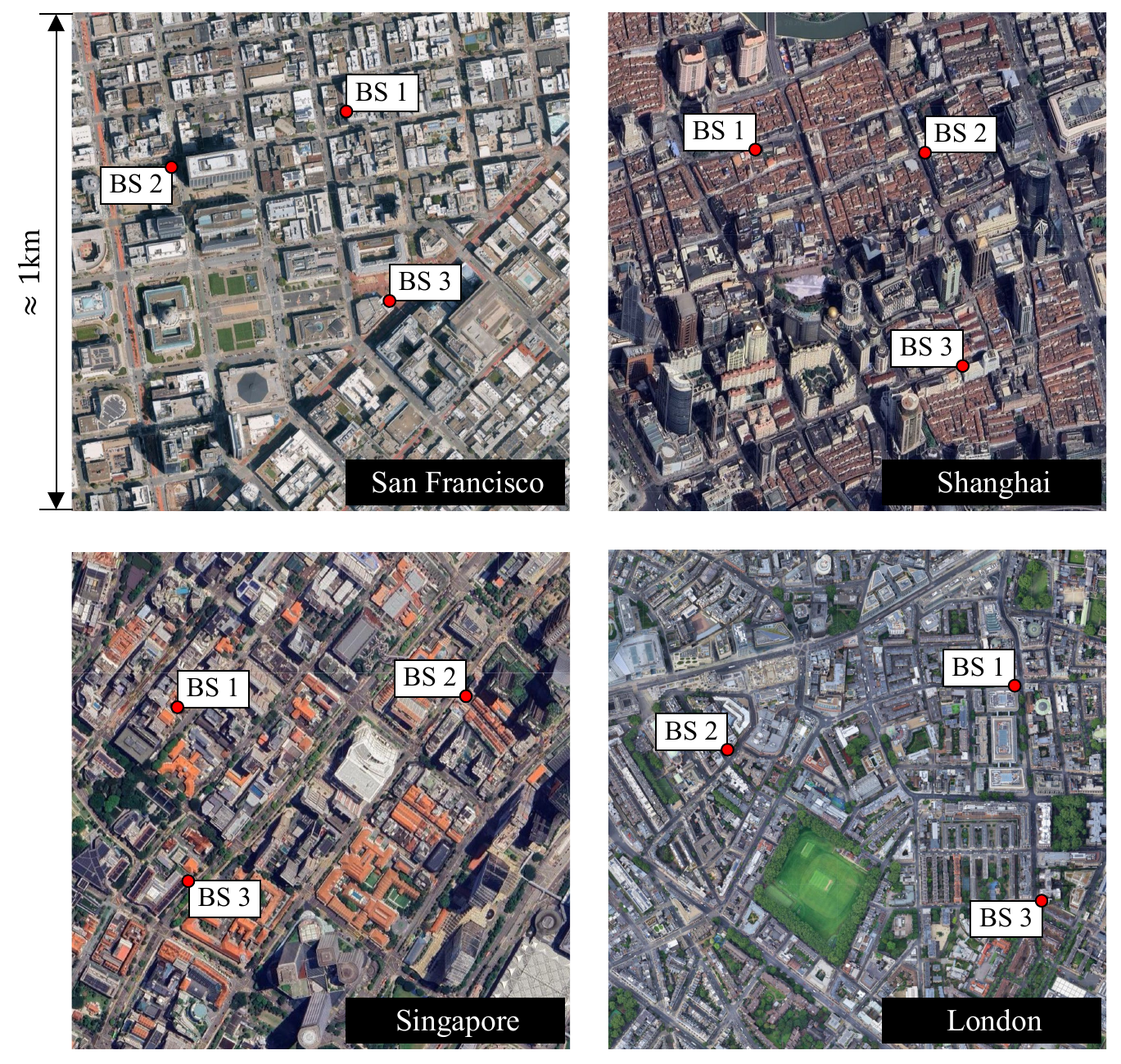}
\caption{Satellite images of the four cities.}
\label{fig::cities}
\vspace{-0.1cm}
\end{figure}

\begin{figure}[t]
\vspace{-0.1cm}
\centering
\includegraphics[width=0.9\columnwidth]{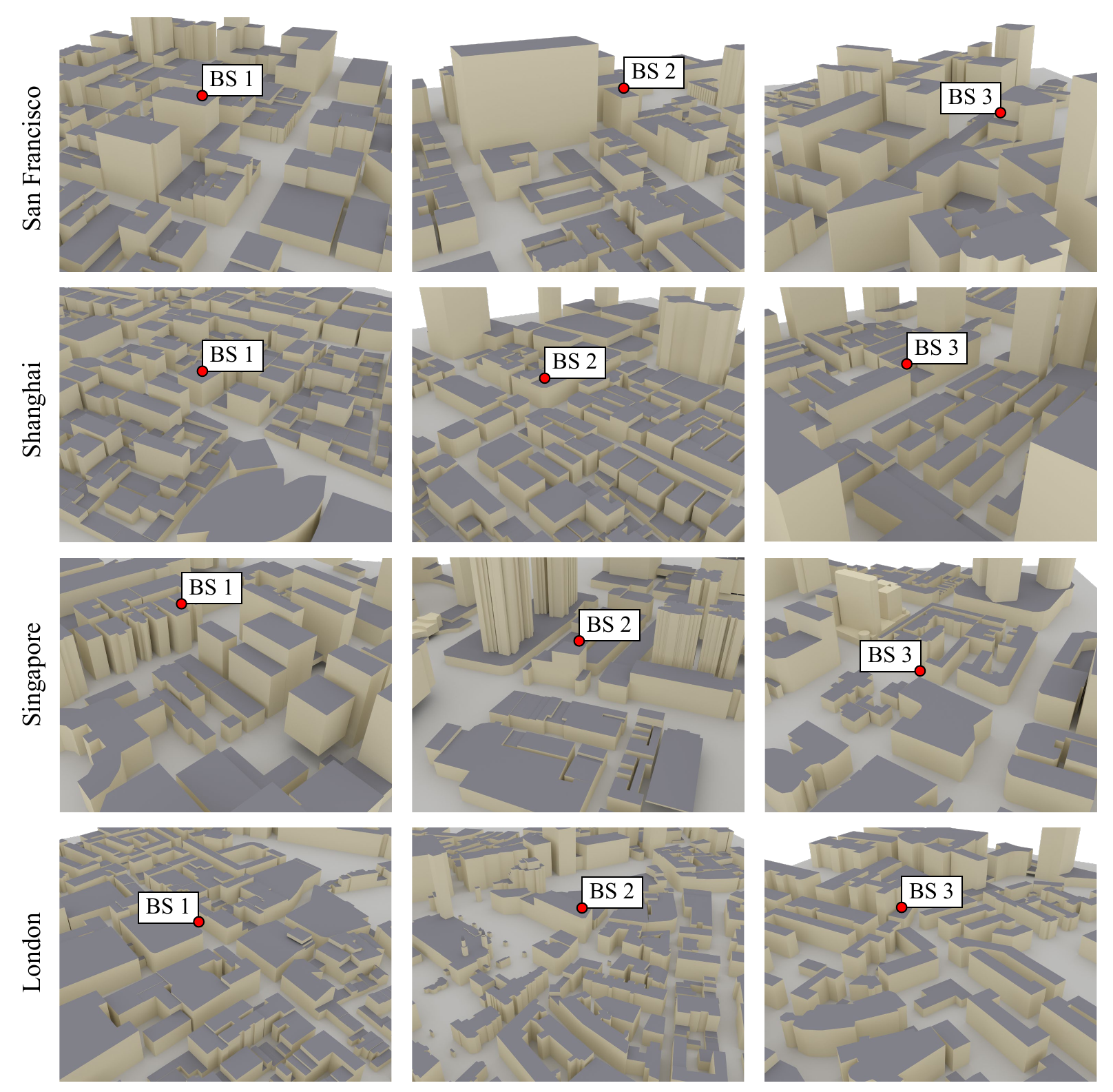}
\caption{Environmental maps of the 12 scenarios.}
\label{fig::scenes}
\vspace{-0.1cm}
\end{figure}

\begin{table}[t]
\centering
\caption{Different settings of pilot patterns.}
\label{tab::pilot_setting}
\begin{tabular}{lll}
\toprule
Setting & Interval & Start \\
\midrule
\multirow{2}{*}{FRFS} & $R_{\rm t}=4$ & $S_{\rm t}=1$ \\
& $R_{\rm c}=16$ & $S_{\rm c}=1$ \\
\midrule
\multirow{2}{*}{FRVS} & $R_{\rm t}=4$ & $S_{\rm t}\in\{1,2,\cdots,R_{\rm t}\}$ \\
& $R_{\rm c}=16$ & $S_{\rm c}\in\{1,2,\cdots,R_{\rm c}\}$ \\
\midrule
\multirow{2}{*}{VRVS} & $R_{\rm t}\in\{2,4,8\}$ & $S_{\rm t}\in\{1,2,\cdots,R_{\rm t}\}$ \\
& $R_{\rm c}\in\{4,8,16,32,64\}$ & $S_{\rm c}\in\{1,2,\cdots,R_{\rm c}\}$ \\
\bottomrule
\end{tabular}
\end{table}

\begin{table}[t]
\centering
\caption{Settings of network structure parameters.}
\label{tab::network_setting}
\begin{tabular}{lll}
\toprule
Scheme & Network Module & Paremeter settings \\
\midrule
\multirow{2}{*}{(m)GCDNet} & (m)CMixer & $K=3$, $N_{\rm t}'=N_{\rm t}$, $N_{\rm c}'=N_{\rm c}$ \\
& Transformer Encoder & $L=6$, $D=512$ \\
\midrule
\multirow{2}{*}{(m)PCMixer} & CMixer & $\widetilde{K}=8$, $N_{\rm t}'=N_{\rm t}$, $N_{\rm c}'=N_{\rm c}$ \\
& ResMLP & $\widetilde{L}=6$, $D=512$ \\
\midrule
(m)CMixer & - & $\widetilde{K}=8$, $N_{\rm t}'=N_{\rm t}$, $N_{\rm c}'=N_{\rm c}$ \\
\bottomrule
\end{tabular}
\end{table}

\begin{table}[t]
\aboverulesep=0pt
\belowrulesep=0pt
\renewcommand{\arraystretch}{1.2}
\centering
\caption{Number of parameters and FLOPs of networks in different schemes, evaluated under $R_{\rm t}=4$, $R_{\rm c}=16$.}
\label{tab::flops}
\begin{tabular}{c|cc|cc}
\toprule
\multirow{2}{*}{Scheme} & \multicolumn{2}{c|}{Mask-free} & \multicolumn{2}{c}{Mask-enhanced} \\
\cmidrule{2-5}
& Parameters & FLOPs & Parameters & FLOPs \\
\midrule
(m)GCDNet & 21.85 M & 610.76 M & 24.73 M & 708.48 M \\
(m)PCMixer & 15.33 M & 182.38 M & 21.56 M & 194.83 M \\
(m)CMixer & 4.51 M & 152.84 M & 10.69 M & 362.21 M \\
(m)PCMixer-L & 43.27 M & 1.25 G & 55.72 M & 1.27 G \\
(m)CMixer-L & 17.44 M & 1.16 G & 40.32 M & 2.69 G \\
\bottomrule
\end{tabular}
\end{table}


\subsection{Experimental Settings}

\subsubsection{Scenario Setup}

We collect the environmental maps of four global cities from OpenStreetMap, including San Francisco, Shanghai, Singapore, and London. For each city, we manually set three BS positions, constructing 12 scenarios. The satellite images of these four cities and the 12 constructed scenarios are shown in Fig. \ref{fig::cities} and Fig. \ref{fig::scenes}, respectively. These environmental maps are imported into Sionna RT \cite{hoydis2023sionnart} to generate channel data and geometric features. The materials of the ground, building walls, and building roofs are set as concrete, marble, and metal, respectively. The propagation paths with no more than 5 reflections are enabled. In each scenario, the users are randomly distributed within a $400\,\text{m}\times 400\,\text{m}$ square area centered on the BS. User heights range from $1\,\text{m}$ to $2\,\text{m}$. Virtual user positions are sampled on a regular cell grid with a cell size of $1\,\text{m}\times 1\,\text{m}$ and a height of $1.5\,\text{m}$. The BS is equipped with a uniform linear array (ULA) with $N_{\rm t}=16$ array elements. All users are equipped with randomly oriented dipole antennas to simulate diverse antenna radiation characteristics. The system center frequency is $f=5\,\text{GHz}$, and the bandwidth is set to 40\,MHz and divided into $N_{\rm c}=256$ subcarriers.

We consider the following pilot pattern settings, with details shown in Table \ref{tab::pilot_setting}. Each setting defines a collection of pilot patterns $\{\Omega_{\ell}\}_{\ell}$ used in training.
\begin{itemize}
    \item Fixed $R$ \& fixed $S$ (FRFS): In each domain (space or frequency), both the pilot interval and start are fixed. In this case, there is only one possible pilot pattern.
    \item Fixed $R$ \& variable $S$ (FRVS): In each domain, the pilot interval is fixed, while the start is variable. In this case, the number of pilots is fixed, but these pilots are inserted at variable positions in the spatial-frequency grid.
    \item Variable $R$ \& variable $S$ (VRVS): In each domain, both the pilot interval and start are variable.
\end{itemize}

\subsubsection{Baselines and Parameter Settings}

We adopt two advanced learning schemes for channel acquisition as baselines. One is CMixer-based channel mapping \cite{chen2024cmixer}, which utilizes a $\widetilde{K}$-layer CMixer to generate the complete channel using only pilot-based estimates. The other is PCMixer \cite{chen2025scd}, which first processes the user position and partial channel separately using two $\widetilde{L}$-layer ResMLPs with hidden size $D$, followed by $1\times 1$ convolutional fusion, and finally generates the full channel via a $\widetilde{K}$-layer CMixer for refinement. The original implementations of CMixer and PCMixer only support fixed pilot pattern. To better align with our work, we extend these two baselines to their mask-enhanced versions. For CMixer-based channel mapping, we introduce mCMixer proposed in the previous section to support channel mapping with variable pilot patterns. For PCMixer, we replace its input $\mathbf{H}_{\pi}$ with the concatenation of $\widetilde{\mathbf{H}}$ and $\mathbf{M}$, yielding mask-enhanced PCMixer (mPCMixer). The detailed parameter settings of these network structures are shown in Table \ref{tab::network_setting}. Besides, we double the hidden dimensions of the network modules in (m)PCMixer and (m)CMixer, resulting in additional baselines named (m)PCMixer-L and (m)CMixer-L, which possess a significantly higher level of complexity than (m)GCDNet.

We employ different numbers of training epochs for the three pilot pattern settings in Table \ref{tab::pilot_setting} due to their difference in training data diversity. Specifically, we train each network for 1000 epochs under the FRFS setting, for 4000 epochs under the FRVS setting, and for 10000 epochs under the VRVS setting. The batch size is set to 500. The learning rate is initially set to $10^{-4}$ and decayed by a factor of 0.8 every 1/10 of the training duration. During training, the number of pseudo channels $n$ is randomly selected from 0 to 16. During testing, $n$ is set to 16 by default. Table \ref{tab::flops} shows the number of parameters and the floating-point operations (FLOPs) of networks in the above schemes\footnote{For GCD, the data volume of the geometric feature set $\mathcal{F}$ (approximately 1\,M) is relatively small compared to the model parameters (approximately 20\,M), and the computational cost for pseudo channel construction (less than 10 MFLOPs) is negligible compared to the neural network inference (hundreds of MFLOPs).}.

In addition, we introduce generative channel estimation as another baseline, which employs diffusion model based posterior sampling (DMPS) \cite{meng2022dmps}. We implement this scheme based on the implementation of \cite{zhou2025generative}, except that we replace the denoising network, originally a CNN, with more advanced diffusion Transformer (DiT) \cite{peebles2023dit}. In our experiment, the DiT consists of 8 DiT blocks with a hidden size of 512 and is trained for 10000 epochs.

To demonstrate the advantages of deep learning methods over traditional signal processing algorithms, we also compare with the classical linear minimum mean squared error (LMMSE) channel estimation, where the covariance matrix is computed based on training data. Furthermore, we incorporate the idea of basis projection proposed in \cite{del2025bayesian} into LMMSE, thereby forming another baseline, LMMSE-BP, which similarly utilizes environmental geometry as prior knowledge. 


\subsubsection{Evaluation Metrics}

We evaluate the channel acquisition quality for each data sample using normalized MSE (NMSE) and cosine correlation $\rho$, which are defined as
\begin{align}
\text{NMSE}&=\|\mathbf{H}-\widehat{\mathbf{H}}\|_{\rm F}^2/\|\mathbf{H}\|_{\rm F}^2,\\
\rho&=\frac{1}{N_{\rm c}}\sum_{n_{\rm c}=1}^{N_{\rm c}}\frac{|\widehat{\bm{h}}_{n_{\rm c}}^{\mathsf{H}}\bm{h}_{n_{\rm c}}|}{\|\widehat{\bm{h}}_{n_{\rm c}}\|_2\|\bm{h}_{n_{\rm c}}\|_2},
\end{align}
where $\bm{h}_{n_{\rm c}},\widehat{\bm{h}}_{n_{\rm c}}$ are the $n_{\rm c}$-th columns of $\mathbf{H},\widehat{\mathbf{H}}$, respectively.

To translate the channel acquisition accuracy into meaningful communication gains, we use the average achievable rate (AAR) to evaluate the performance of precoding based on the acquired CSI, which is defined as
\begin{equation}
\text{AAR}=\frac{1}{N_{\rm c}}\sum_{n_{\rm c}=1}^{N_{\rm c}}\log_2\left(1+\frac{P}{\sigma_{\rm w}^2}|\bm{h}_{n_{\rm c}}^{\mathsf{T}}\bm{v}_{n_{\rm c}}|^2\right)(\text{bps/Hz}),
\end{equation}
where $\bm{v}_{n_{\rm c}}\in\mathbb{C}^{N_{\rm t}}$ denotes the precoding vector for the $n_{\rm c}$-th subcarrier, which is assumed to have a unit power, i.e., $\|\bm{v}_{n_{\rm c}}\|_2^2=1$. For simplicity, we employ maximum ratio transmission (MRT), whose precoding vector is given by $\bm{v}_{n_{\rm c}}=\widehat{\bm{h}}_{n_{\rm c}}^*/\|\widehat{\bm{h}}_{n_{\rm c}}\|_2$, where the acquired channel $\widehat{\bm{h}}_{n_{\rm c}}$ accounts for estimation noise (with a power of $\widetilde{\sigma}_{\rm w}^2=\sigma_{\rm w}^2/P$). As a reminder, $P$ and $\sigma_{\rm w}^2$ represent the BS transmit power and thermal noise power, respectively. The thermal noise power can be calculated as $\sigma_{\rm w}^2=kTB$, where $k=1.38\times 10^{-23}\,\text{J/K}$ is the Boltzmann constant, $T=290\,\text{K}$ is the temperature, and $B$ is the system bandwidth.

\begin{figure*}[t]
\vspace{-0.1cm}
\centering
\includegraphics[width=2\columnwidth]{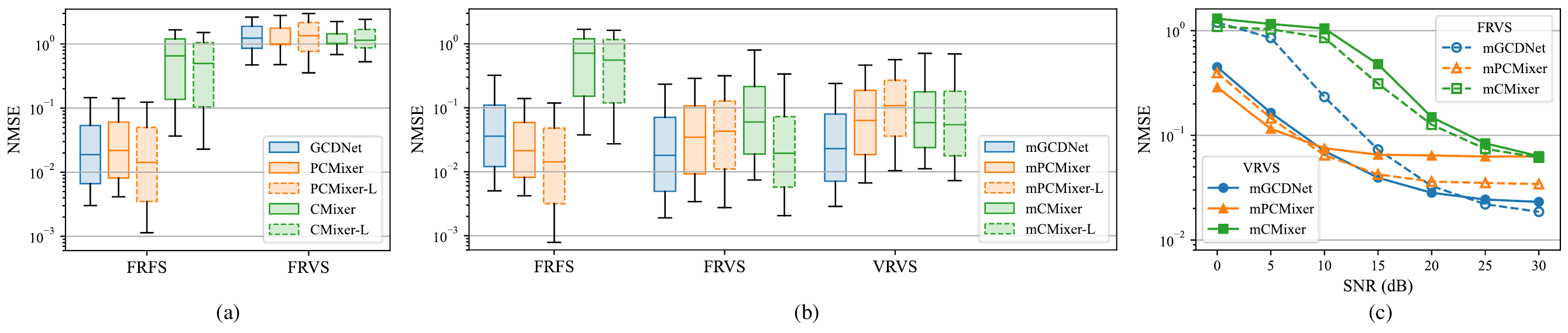}
\caption{NMSE performance with networks trained under different pilot pattern settings. In the box plots, the boxes extend from the first quartile (Q1) to the third quartile (Q3) with a line at the median, and the whiskers extend from the 10th percentile to the 90th percentile. (a) NMSE box plot of mask-free schemes. (b) NMSE box plot of mask-enhanced schemes. (c) Median NMSE of mask-enhanced schemes under different noise conditions.}
\label{fig::results_pilot_setting}
\vspace{-0.1cm}
\end{figure*}

\begin{table*}[t]
\aboverulesep=0pt
\belowrulesep=0pt
\renewcommand{\arraystretch}{1.2}
\centering
\caption{NMSE and $\rho$ performance with networks trained under different pilot pattern settings.}
\label{tab::results_pilot_setting}
\begin{tabular}{c|cc|cc|cc|cc|cc|cc}
\toprule
\multirow{3}{*}{Scheme} & \multicolumn{4}{c|}{FRFS} & \multicolumn{4}{c|}{FRVS} & \multicolumn{4}{c}{VRVS} \\
\cmidrule{2-13}
& \multicolumn{2}{c|}{NMSE (dB)} & \multicolumn{2}{c|}{$\rho$} & \multicolumn{2}{c|}{NMSE (dB)} & \multicolumn{2}{c|}{$\rho$} & \multicolumn{2}{c|}{NMSE (dB)} & \multicolumn{2}{c}{$\rho$} \\
& Mean & Median & Mean & Median & Mean & Median & Mean & Median & Mean & Median & Mean & Median \\
\midrule
mGCDNet & -8.60 & -14.45 & 0.9412 & 0.9854 & \textbf{-9.95} & \textbf{-17.46} & \textbf{0.9614} & \textbf{0.9940} & \textbf{-9.69} & \textbf{-16.37} & \textbf{0.9602} & \textbf{0.9924} \\
mPCMixer & -11.98 & -16.65 & 0.9712 & 0.9903 & -9.29 & -14.68 & 0.9550 & 0.9876 & -7.53 & -12.01 & 0.9359 & 0.9801 \\
mCMixer & -1.17 & -1.47 & 0.6001 & 0.7424 & -6.18 & -12.24 & 0.9034 & 0.9805 & -6.77 & -12.45 & 0.9061 & 0.9825 \\
mPCMixer-L & \textbf{-12.97} & \textbf{-18.47} & \textbf{0.9782} & \textbf{0.9944} & -8.85 & -13.67 & 0.9518 & 0.9857 & -6.44 & -9.67 & 0.9264 & 0.9700 \\
mCMixer-L & -1.58 & -2.55 & 0.6440 & 0.8134 & -8.61 & -17.12 & 0.9410 & 0.9937 & -7.02 & -12.65 & 0.9122 & 0.9852 \\
\bottomrule
\end{tabular}
\end{table*}

\subsection{Single-Scenario Learning}

In this subsection, we train and evaluate all schemes in a single scenario -- San Francisco BS 1. The training, validation, and testing datasets comprise 40\,k, 10\,k, and 10\,k data samples, respectively.

\subsubsection{Training under Different Pilot Pattern Settings}

To verify the effectiveness of the mask enhancement, we evaluate the channel acquisition performance with networks trained under different pilot pattern settings, i.e., FRFS, FRVS, and VRVS. For fair comparison, all schemes are evaluated under $R_{\rm t}=4$, $R_{\rm c}=16$. The NMSE performance of mask-free schemes are shown in Fig. \ref{fig::results_pilot_setting}(a). When trained and evaluated with fixed pilot pattern (FRFS), CMixer struggles to perform effectively due to the complex multi-path characteristics and limited pilot resources. In contrast, both PCMixer and GCDNet utilize additional information, thereby achieving superior accuracy.
However, under the FRVS setting, all these schemes fail because they do not utilize the necessary information about pilot patterns. Worse still, these schemes are not applicable to the VRVS setting, as their network structures strictly determine the partial channel size, thus only supporting fixed pilot interval.

The NMSE performance of mask-enhanced schemes are shown in Fig. \ref{fig::results_pilot_setting}(b). All schemes perform well under the FRVS and VRVS settings. Notably, the performance of mGCDNet and mCMixer under the FRVS and VRVS settings is even better than under the FRFS setting, indicating that variable pilot patterns enrich the training data and facilitate more comprehensive learning. Furthermore, it can be observed that under the FRVS and VRVS settings, the proposed mGCDNet achieves the best performance among all schemes, even though it has much fewer parameters and FLOPs than mPCMixer-L or mCMixer-L. Table \ref{tab::results_pilot_setting} presents the quantitative results under different settings, further validating the superiority of the proposed mGCDNet under variable pilot configurations.

In Fig. \ref{fig::results_pilot_setting}(c), we evaluate mask-enhanced schemes with variable pilot patterns under different noise conditions. The noisy estimate can be expressed as $\mathbf{H}_{\pi}'=\mathbf{H}_{\pi}+\mathbf{W}$, where $\mathbf{W}\in\mathbb{C}^{N_{\rm t,\pi}\times N_{\rm c,\pi}}$ is the additive noise matrix whose elements are independently sampled from $\mathcal{CN}(0,\widetilde{\sigma}_{\rm w}^2)$, with $\widetilde{\sigma}_{\rm w}^2$ being the estimation noise power. We define the signal-to-noise ratio (SNR) as $P_{\mathbf{H}}/\widetilde{\sigma}_{\rm w}^2$. As SNR decreases, the performance of all schemes degrades. Notably, mGCDNet exhibits stronger robustness to input noise under VRVS compared to FRVS, which once again validates the advantages of enriching training pilot configurations.

In the remainder of this section, we evaluate the mask-enhanced schemes that are trained under the VRVS setting, as it improves performance, robustness, and universality.

\begin{figure*}[t]
\vspace{-0.1cm}
\centering
\includegraphics[width=1.9\columnwidth]{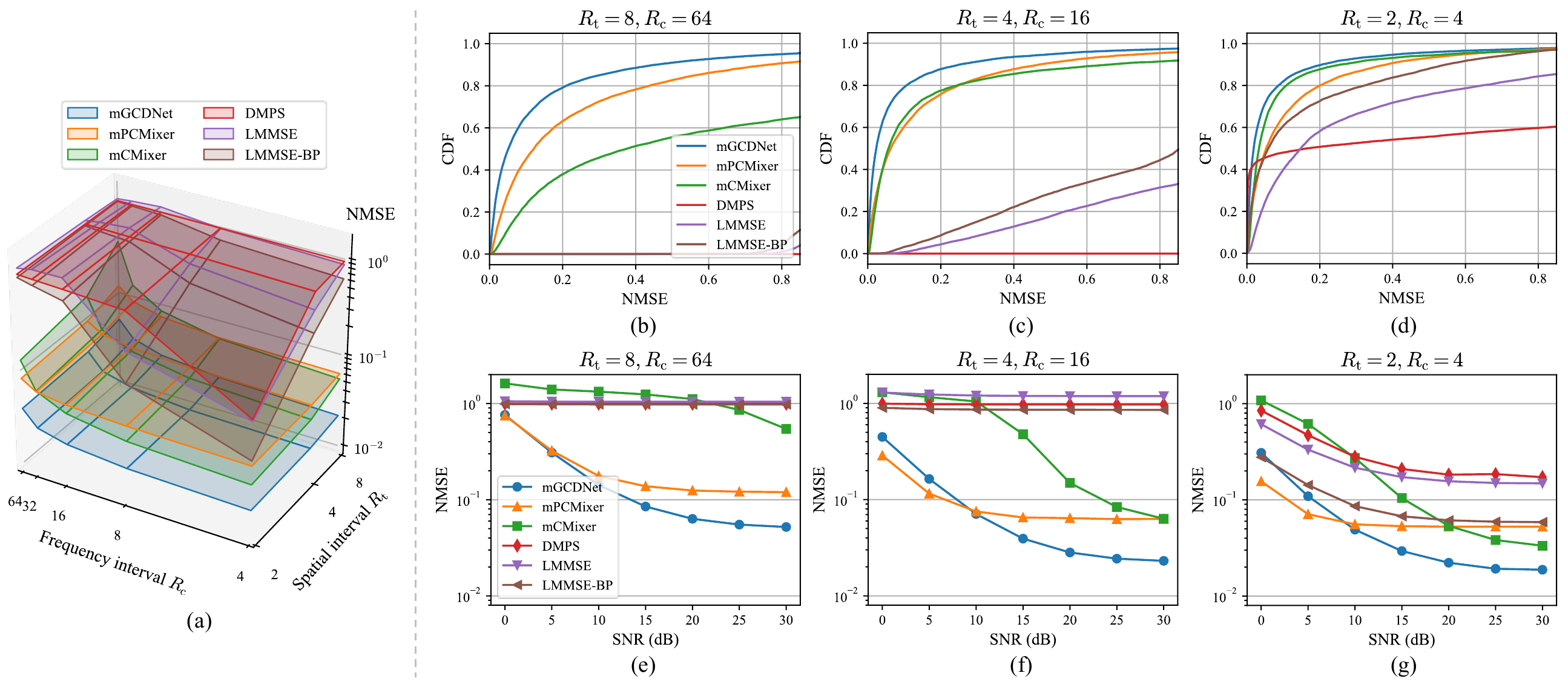}
\caption{NMSE performance under different pilot intervals. (a) Median NMSE under various groups of $(R_{\rm t},R_{\rm c})$. (b)-(d) Cumulative probability distribution of NMSE with $(R_{\rm t},R_{\rm c})\in\{(8,64),(4,16),(2,4)\}$. (e)-(g) Median NMSE under different noise conditions with $(R_{\rm t},R_{\rm c})\in\{(8,64),(4,16),(2,4)\}$.}
\label{fig::results_ratio}
\vspace{-0.1cm}
\end{figure*}

\begin{table*}[t]
\aboverulesep=0pt
\belowrulesep=0pt
\renewcommand{\arraystretch}{1.2}
\centering
\caption{NMSE and $\rho$ performance under various pilot intervals.}
\label{tab::results_ratio}
\begin{tabular}{c|cc|cc|cc|cc|cc|cc}
\toprule
\multirow{3}{*}{Scheme} & \multicolumn{4}{c|}{$R_{\rm t}=8$, $R_{\rm c}=64$} & \multicolumn{4}{c|}{$R_{\rm t}=4$, $R_{\rm c}=16$} & \multicolumn{4}{c}{$R_{\rm t}=2$, $R_{\rm c}=4$} \\
\cmidrule{2-13}
& \multicolumn{2}{c|}{NMSE (dB)} & \multicolumn{2}{c|}{$\rho$} & \multicolumn{2}{c|}{NMSE (dB)} & \multicolumn{2}{c|}{$\rho$} & \multicolumn{2}{c|}{NMSE (dB)} & \multicolumn{2}{c}{$\rho$} \\
& Mean & Median & Mean & Median & Mean & Median & Mean & Median & Mean & Median & Mean & Median \\
\midrule
mGCDNet & \textbf{-7.70} & \textbf{-12.90} & \textbf{0.9437} & \textbf{0.9857} & \textbf{-9.69} & \textbf{-16.37} & \textbf{0.9602} & \textbf{0.9924} & \textbf{-10.43} & \textbf{-17.36} & \textbf{0.9643} & 0.9938 \\
mPCMixer & -5.59 & -9.18 & 0.9143 & 0.9694 & -7.53 & -12.01 & 0.9359 & 0.9801 & -8.48 & -12.79 & 0.9424 & 0.9819 \\
mCMixer & -1.91 & -4.28 & 0.6987 & 0.8982 & -6.77 & -12.45 & 0.9061 & 0.9825 & -9.47 & -14.82 & 0.9492 & 0.9890 \\
DMPS & -0.01 & -0.01 & 0.2418 & 0.1584 & -0.12 & -0.12 & 0.3619 & 0.3499 & -1.32 & -7.87 & 0.6611 & 0.9518 \\
LMMSE & 0.07 & 0.18 & 0.3106 & 0.3176 & 0.05 & 0.75 & 0.4773 & 0.4814 & -4.59 & -8.32 & 0.8187 & 0.9253 \\
LMMSE-BP & -0.23 & -0.10 & 0.8626 & 0.9259 & -1.52 & -0.68 & 0.9163 & 0.9664 & -7.65 & -12.33 & 0.9593 & \textbf{0.9947} \\
\bottomrule
\end{tabular}
\end{table*}

\subsubsection{Performance under Different Pilot Intervals}

Fig. \ref{fig::results_ratio}(a) illustrates the performance of different schemes under various pilot intervals $(R_{\rm t},R_{\rm c})$. To provide a more comprehensive demonstration, we select three groups of pilot intervals $(R_{\rm t},R_{\rm c})$, plotting the cumulative distribution function (CDF) of NMSE under the ideal noise-free condition in Fig. \ref{fig::results_ratio}(b)-(d) and the NMSE performance under different SNR in Fig. \ref{fig::results_ratio}(e)-(g). Table \ref{tab::results_ratio} shows the detailed quantitative results under the noise-free condition. Clearly, DMPS and LMMSE can work effectively only when pilot resources are abundant (e.g., $R_{\rm t}=2$, $R_{\rm c}=4$), because when only a small number of pilots are observed, neither the diffusion model nor the covariance matrix provides sufficient prior information to determine the complete CSI.
Although LMMSE-BP improves LMMSE's performance by utilizing basis projection based on geometric prior, it relies heavily on the initial performance of LMMSE channel estimation. Therefore, poor LMMSE performance hinders the performance improvement of LMMSE-BP. mCMixer also exhibits severe performance degradation under conditions with very few pilots (e.g., $R_{\rm t}=8$, $R_{\rm c}=64$). In contrast, mPCMixer and mGCDNet operate properly across various pilot intervals by utilizing auxiliary information, and mGCDNet consistently outperforms the other schemes.

In the remainder of this section, all schemes are evaluated under $R_{\rm t}=4$, $R_{\rm c}=16$ to strike a balance between channel acquisition quality and pilot overhead.

\begin{figure}[t]
\vspace{-0.1cm}
\centering
\includegraphics[width=\columnwidth]{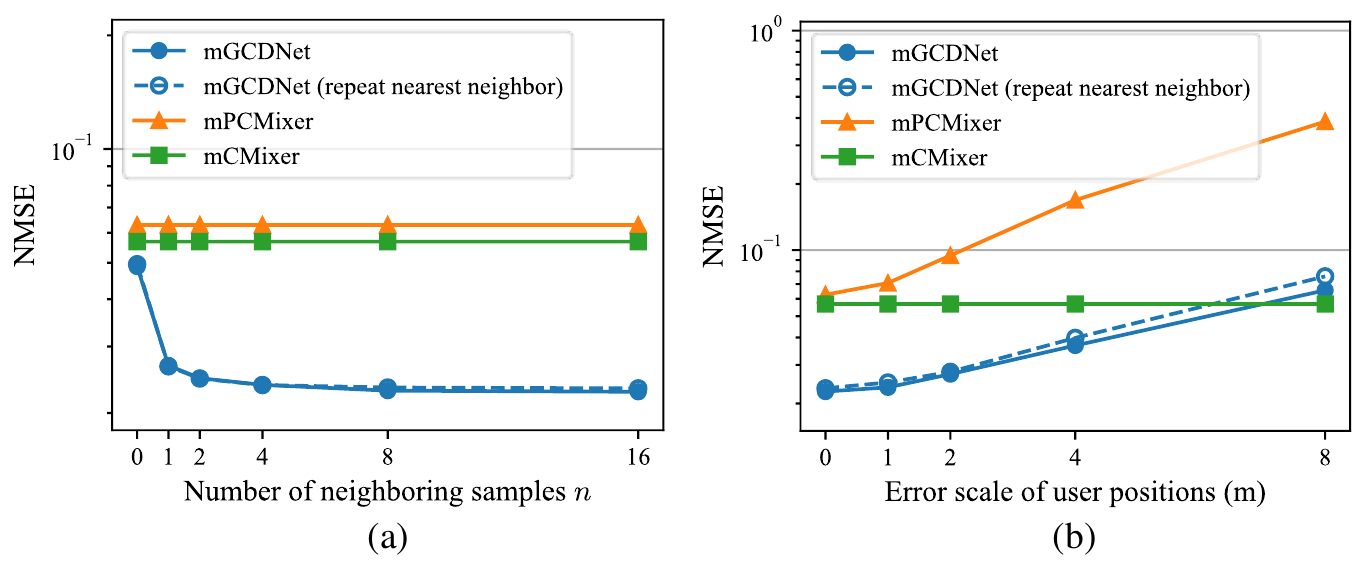}
\caption{Impact of prompt length (i.e., the number of neighboring samples) and position error. (a) Median NMSE under different numbers of neighboring samples. (b) Median NMSE under different scales of user position errors.}
\label{fig::results_sf1}
\vspace{-0.1cm}
\end{figure}

\subsubsection{Impact of Prompt Length and Position Error}

Fig. \ref{fig::results_sf1}(a) illustrates mGCDNet’s performance under various prompt lengths, i.e., the number of neighboring samples $n$. As $n$ increases, the performance of mGCDNet first improves and then tends to plateau. Even with only a single neighboring sample, mGCDNet still outperforms other schemes by a wide margin. Recall that the pseudo channels $\widetilde{\mathbf{H}}_i(i=i_1,\cdots,i_n)$ are derived from the geometric features of $n$ neighboring samples in $\mathcal{F}$, with each sample assigned a random placeholder $\widetilde{\bm{z}}_i$. Here, we try a different strategy: instead of searching for $n$ neighboring samples, we use only the nearest one and repeat its geometric features for $n$ times, assigning a different placeholder each time. The results are shown by the dashed lines in Fig. \ref{fig::results_sf1}(a). We observe that the performance difference between reusing a single neighbor $n$ times and using $n$ neighbors is negligible. This suggests that the performance gain from increasing $n$ primarily stems from the multiple realizations of the random placeholder.

Next, we introduce errors to user positions. The approximate position is represented as $\widehat{\bm{x}}=\bm{x}+\Delta\bm{x}$, where $\Delta\bm{x}$ is the position error following a 2D uniform distribution on $[-l,l]\times[-l,l]$, with $l$ being the error scale\footnote{Here, we represent the user position as a 2D vector, omitting height information. Since user heights typically fall within in a narrow range (e.g., $[1\,\text{m},2\,\text{m}]$), we simply assign all virtual user positions the same height (1.5\,m). Consequently, real user height values are unnecessary here as they do not affect the neighborhood search result.}. The results are shown in Fig. \ref{fig::results_sf1}(b). As the position error increases, the performance of both mGCDNet and mPCMixer degrades. When $l<4\,\text{m}$, mGCDNet consistently demonstrates a performance advantage over the other schemes, performing well within the positioning accuracy achievable by existing systems. We also observe that using $n$ distinct neighboring samples exhibits slightly stronger robustness against position errors than reusing the nearest neighbor $n$ times. This is because, even if the position used for neighborhood sampling is significantly biased, multiple neighboring samples may still form a region that covers the true position and provides approximate structural features \cite{chen2025al}.


\begin{figure}[t]
\vspace{-0.1cm}
\centering
\includegraphics[width=0.95\columnwidth]{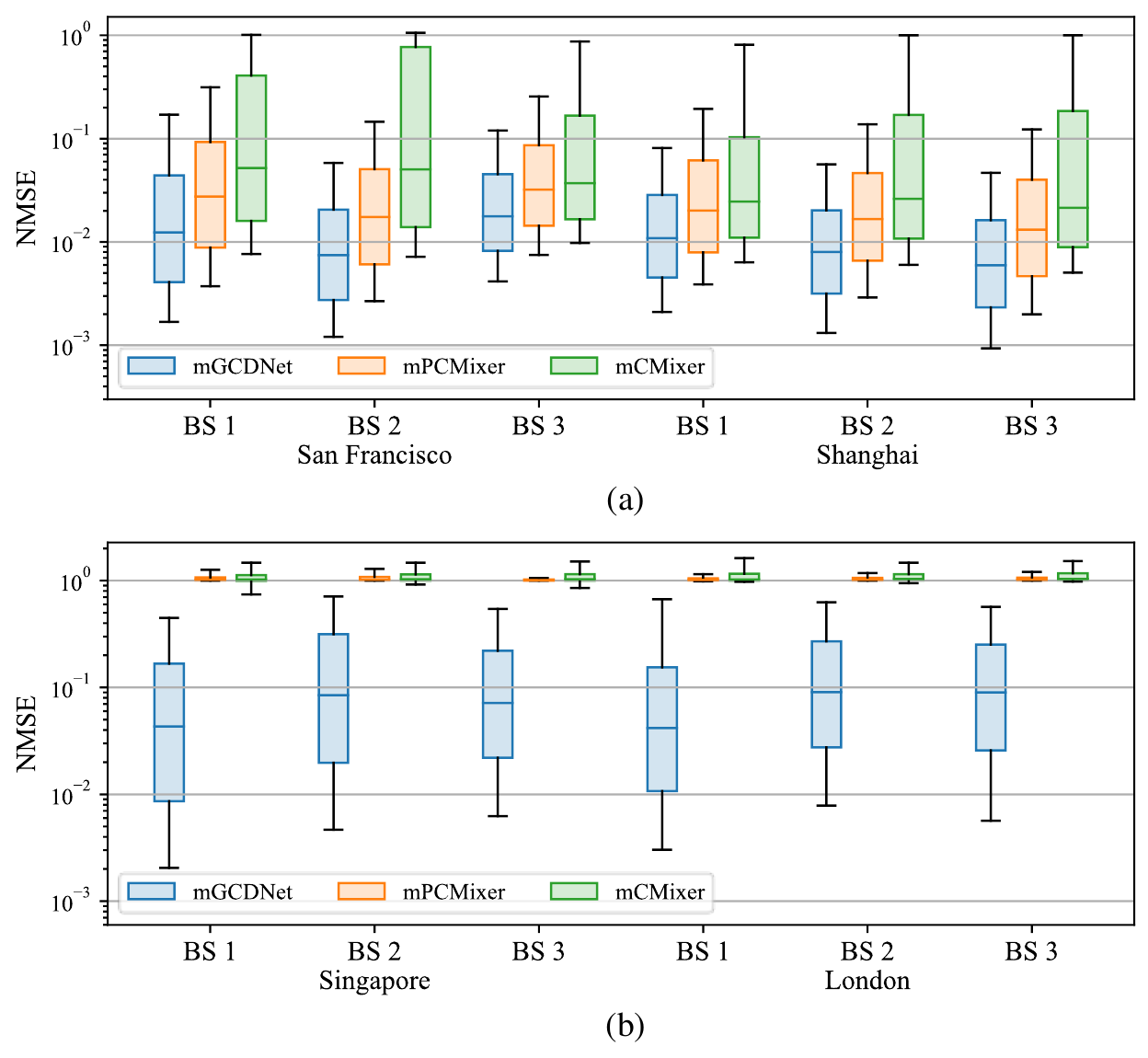}
\caption{NMSE performance in multi-scenario learning. (a) NMSE box plots in learned scenarios. (b) NMSE box plots in unseen scenarios without additional training.}
\label{fig::results_all}
\vspace{-0.1cm}
\end{figure}

\begin{table}[t]
\aboverulesep=0pt
\belowrulesep=0pt
\renewcommand{\arraystretch}{1.2}
\centering
\caption{NMSE and $\rho$ performance in two example scenarios.}
\label{tab::results_sfsg}
\begin{tabular}{c|c|cc|cc}
\toprule
\multirow{2}{*}{Scenario} & \multirow{2}{*}{Scheme} & \multicolumn{2}{c|}{NMSE (dB)} & \multicolumn{2}{c}{$\rho$} \\
& & Mean & Median & Mean & Median \\
\midrule
\multirow{3}{*}{\makecell[c]{San Francisco\\BS 1}}
& mGCDNet & \textbf{-11.48} & \textbf{-19.08} & \textbf{0.9678} & \textbf{0.9955} \\
& mPCMixer & -9.59 & -15.61 & 0.9515 & 0.9908 \\
& mCMixer & -5.29 & -12.84 & 0.8505 & 0.9820 \\
\midrule
\multirow{3}{*}{\makecell[c]{Singapore\\BS 1}}
& mGCDNet & \textbf{-8.28} & \textbf{-13.64} & \textbf{0.9298} & \textbf{0.9859} \\
& mPCMixer & 0.31 & 0.09 & 0.1828 & 0.0944 \\
& mCMixer & 0.24 & 0.11 & 0.3519 & 0.2238 \\
\bottomrule
\end{tabular}
\end{table}

\begin{figure}[t]
\vspace{-0.1cm}
\centering
\includegraphics[width=\columnwidth]{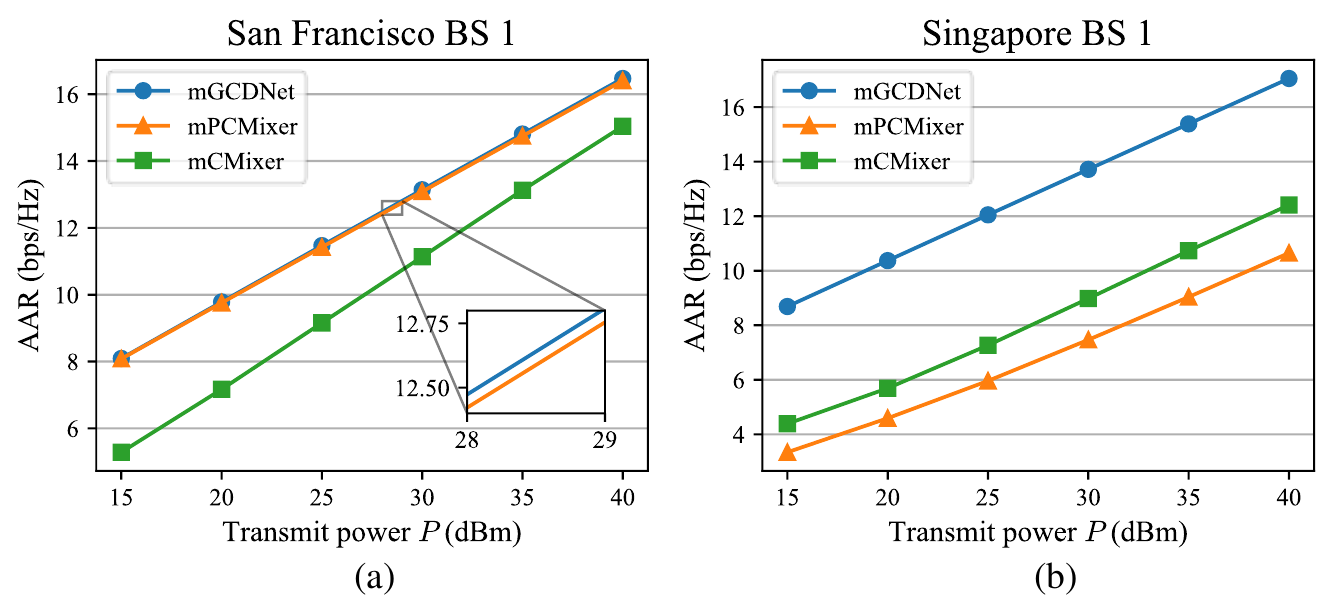}
\caption{AAR performance under different BS transmit power values. (a) Mean AAR in a learned scenario. (b) Mean AAR in an unseen scenario.}
\label{fig::results_aar}
\vspace{-0.1cm}
\end{figure}

\subsection{Multi-Scenario Learning}

In this subsection, we train each scheme using training data from the first 6 scenarios jointly and evaluate it across all 12 scenarios. The training, validation, and testing datasets for each scenario contain 20\,k, 10\,k, and 10\,k data samples, respectively.

\subsubsection{Performance in Original and New Scenarios}

Fig. \ref{fig::results_all}(a) shows the NMSE performance of different schemes across the first 6 scenarios involved in training. Both mGCDNet and mPCMixer utilize auxiliary information to enhance channel acquisition accuracy and achieve multi-scenario joint learning, with mGCDNet consistently outperforming mPCMixer. Fig. \ref{fig::results_all}(b) shows the NMSE performance across the remaining 6 scenarios, where all schemes are evaluated directly without additional training or finetuning. mPCMixer and mCMixer completely fail in these unseen scenarios. In contrast, mGCDNet demonstrates excellent generalization ability, achieving satisfactory NMSE performance on most testing samples in the new scenarios. Table \ref{tab::results_sfsg} presents the quantitative results of different schemes in San Francisco BS 1 and Singapore BS 1, which serve as examples of the learned scenarios and the unseen scenarios, respectively. These results further validate the superiority of the proposed mGCDNet.

Besides, we introduce AAR as an additional evaluation metric. Fig. \ref{fig::results_aar} illustrates the precoding performance under different BS transmit power values $P$ in the two example scenarios. mGCDNet exhibits particular advantage in unseen scenarios, with AAR values comparable to those in learned scenarios and significantly outperforming other schemes. This result demonstrates the feasibility of mGCDNet for precoding in cross-scenario applications.


\begin{figure*}[t]
\vspace{-0.1cm}
\centering
\includegraphics[width=1.9\columnwidth]{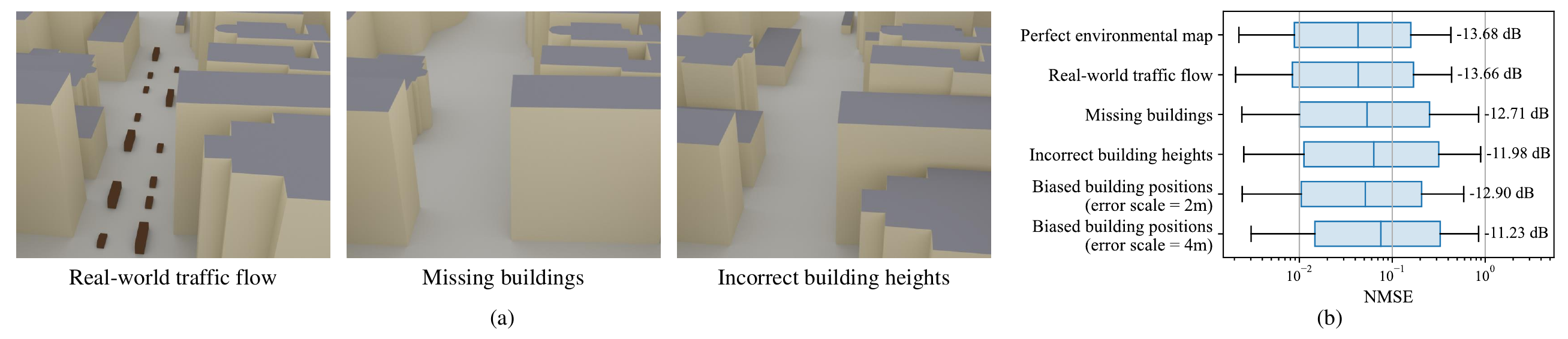}
\caption{Experiments on non-ideal environmental geometry. (a) Example cases of the environmental map. (b) NMSE box plot of mGCDNet in various cases. }
\label{fig::results_biased_geometry}
\vspace{-0.1cm}
\end{figure*}

\subsubsection{Performance under Non-Ideal Geometric Information}

To verify mGCDNet's robustness against non-ideal geometric information, we directly test mGCDNet in one of the unseen scenarios -- Singapore BS 1. First, we consider the following cases of inaccurate environmental maps:
\begin{itemize}
    \item \textbf{Real-world traffic flow:} There are moving vehicles that affect the channel, but the environmental map captures only static buildings. In our experiment, we manually place several vehicles in the environmental map, as shown in Fig. \ref{fig::results_biased_geometry}(a) (left). We then regenerate the channel data while keeping the geometric features unchanged.
    \item \textbf{Missing buildings:} Some buildings are missing from the environmental map. We manually remove some buildings near the base station to ensure a significant change in the propagation paths, as shown in Fig. \ref{fig::results_biased_geometry}(a) (middle), and then regenerate the geometric features.
    \item \textbf{Incorrect building heights:} Some buildings in the environmental map lack accurate height information. In this case, we manually adjust the heights of some buildings, as shown in Fig. \ref{fig::results_biased_geometry}(a) (right), and then regenerate the geometric features.
    \item \textbf{Building position errors:} The building positions in the environmental map are biased. Similar to inaccurate user positions, we independently introduce biased positions for each building in the environmental map. We then regenerate the geometric features.
\end{itemize}

The results are shown in Fig. \ref{fig::results_biased_geometry}(b). Under traffic conditions, mGCDNet exhibits little performance change. This is because, compared to static background buildings, moving vehicles are relatively small in size and thus have a relatively minor impact on the channel. In contrast, non-ideal information regarding the buildings has a greater impact on mGCDNet’s performance, as these cases may alter the number or visibility of propagation paths. Nevertheless, mGCDNet maintains satisfactory performance on most testing samples.

\begin{figure}[t]
\vspace{-0.1cm}
\centering
\includegraphics[width=\columnwidth]{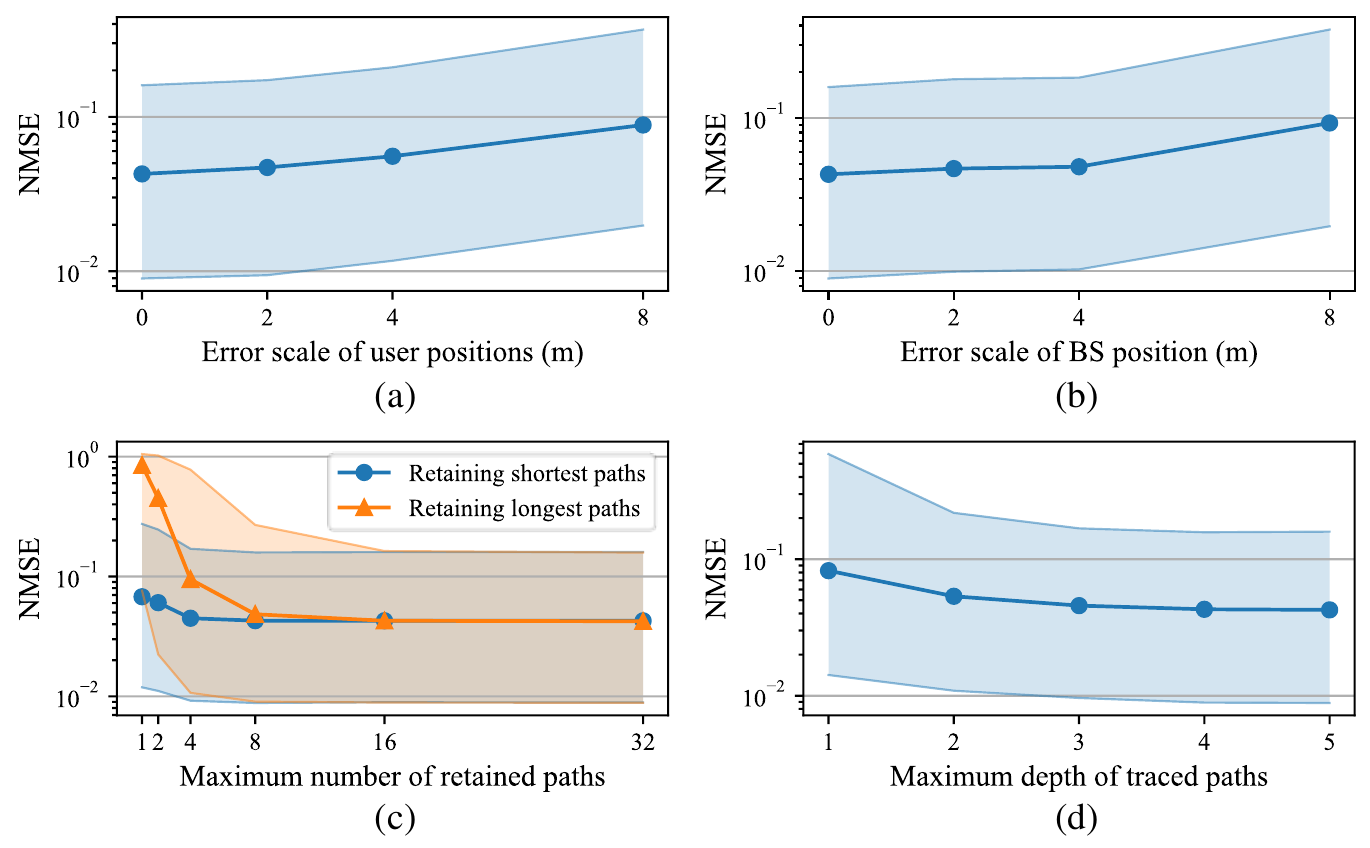}
\caption{NMSE performance of mGCDNet under BS/user position errors and imperfect ray tracing results, where the median NMSE and quartiles are presented. (a)-(b) NMSE under different scales of user/BS position errors. (c) NMSE under different limits of path number. (d) NMSE under different limits of path depths. }
\label{fig::results_pos_rt_error}
\vspace{-0.1cm}
\end{figure}

In Fig. \ref{fig::results_pos_rt_error}(a)-(b), we introduce errors to the BS position and user positions. It can be observed that the performance exhibits a similar trend as the BS/user position error increases.

Next, we consider the imperfect ray tracing results. In Fig. \ref{fig::results_pos_rt_error}(c), we evaluate the impact of the number of paths by limiting the maximum number and removing exceeding paths. As the number of retained paths decreases, the NMSE performance of mGCDNet gradually declines. Meanwhile, the NMSE performance also depends on the lengths of the retained paths, as shorter paths usually exhibit less attenuation and have a greater impact on the channel. Hence, retaining shorter paths results in a smaller performance decline compared to retaining the same number of longer paths. In Fig. \ref{fig::results_pos_rt_error}(d), we evaluate the impact of path depth. i.e., the number of interactions with the environment. As the maximum depth decreases, the NMSE performance of mGCDNet also degrades. Overall, mGCDNet maintains stable performance when the number of retained paths exceeds 8 to 16 and the maximum depth is at least 3.

\begin{figure}[t]
\vspace{-0.1cm}
\centering
\includegraphics[width=\columnwidth]{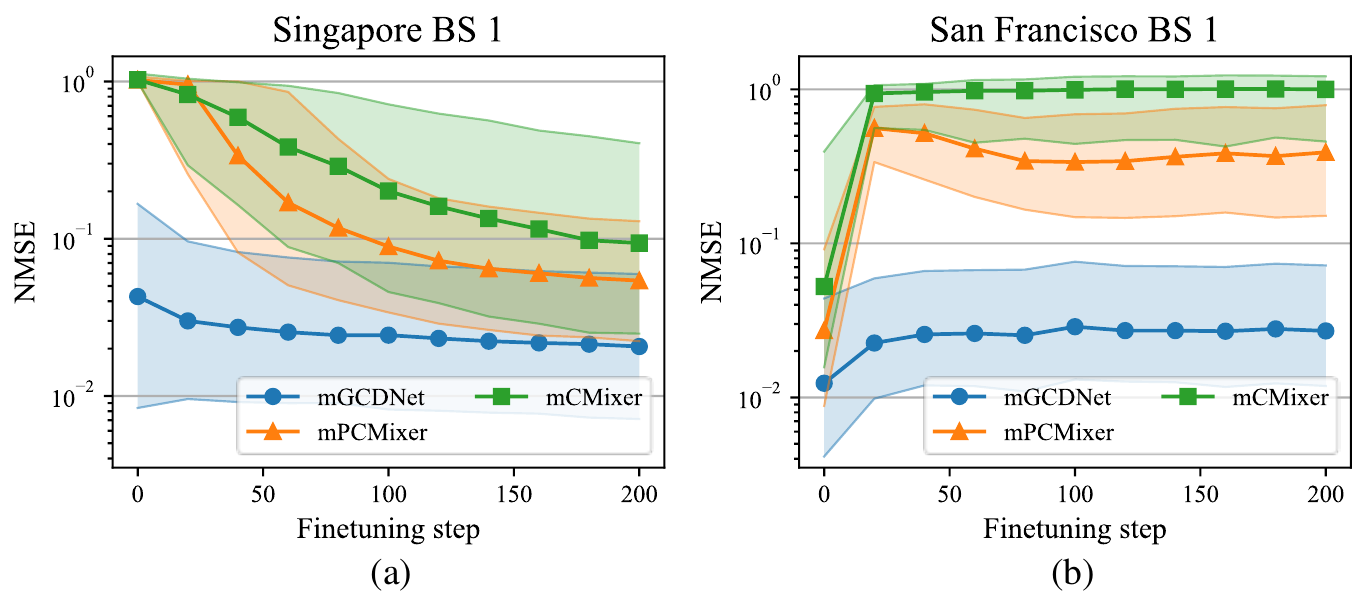}
\caption{NMSE performance after finetuning in the new scenario Singapore BS 1. (a) NMSE in the new scenario. (b) NMSE in a previous scenario.}
\label{fig::results_finetune}
\vspace{-0.1cm}
\end{figure}

\subsubsection{Finetuning in New Scenarios}

When deployed in unseen scenarios, performance degradation on hard testing samples is inevitable due to the significant difference in data distribution compared to the previous scenarios involved in training. Nevertheless, by leveraging the common knowledge gained from pre-training in previous scenarios, mGCDNet can efficiently adapt to these new scenarios with only a small amount of training data and computational resources. To illustrate this point, we perform a few steps of finetuning using data from the new scenario Singapore BS 1. As shown in Fig. \ref{fig::results_finetune}(a), even after hundreds of finetuning steps, the performance of mCMixer and mPCMixer in the new scenario remains inferior to the initial performance of mGCDNet without any finetuning. Furthermore, as shown in Fig. \ref{fig::results_finetune}(b), both mCMixer and mPCMixer exhibit catastrophic forgetting in previous scenarios, resulting in severe performance degradation. In contrast, mGCDNet benefits from multi-scenario pre-training, enabling it to quickly adapt to new scenarios while maintaining outstanding performance in previous ones. More specifically, with only a small number of finetuning steps, mGCDNet can rapidly generalize well to those hard samples in new scenarios, thereby achieving more concentrated NMSE distribution.

\begin{table}[t]
\aboverulesep=0pt
\belowrulesep=0pt
\renewcommand{\arraystretch}{1.2}
\setlength\tabcolsep{5.6pt}
\centering
\caption{System configurations of different scenarios. }
\label{tab::dataset_x}
\begin{tabular}{cc|cccccc}
\toprule
\multicolumn{2}{c|}{Scenario} & Frequency & Bandwidth & \multicolumn{2}{c}{BS antenna} \\
\midrule
\multirow{3}{*}{\makecell[c]{San\\Francisco}}
& BS 1 & 5 GHz & 40 MHz & $1\times 16$ & Short dipole \\
& BS 2 & 3.5 GHz & 20 MHz & $1\times 16$ & Short dipole \\
& BS 3 & 5 GHz & 20 MHz & $4\times 4$ & $\lambda/2$ dipole \\
\midrule
\multirow{3}{*}{Shanghai}
& BS 1 & 6.7 GHz & 50 MHz & $2\times 8$ & Isotropic \\
& BS 2 & 2.4 GHz & 20 MHz & $2\times 8$ & TR 38.901 \\
& BS 3 & 5.9 GHz & 46 MHz & $4\times 4$ & TR 38.901 \\
\midrule
\multirow{3}{*}{Singapore}
& BS 1 & 3.5 GHz & 20 MHz & $2\times 8$ & $\lambda/2$ dipole \\
& BS 2 & 5 GHz & 34 MHz & $4\times 4$ & $\lambda/2$ dipole \\
& BS 3 & 6.7 GHz & 40 MHz & $1\times 16$ & TR 38.901 \\
\midrule
\multirow{3}{*}{London}
& BS 1 & 4.6 GHz & 20 MHz & $4\times 4$ & TR 38.901 \\
& BS 2 & 5 GHz & 40 MHz & $4\times 4$ & Isotropic \\
& BS 3 & 2.4 GHz & 20 MHz & $1\times 16$ & Short dipole \\
\midrule
\multicolumn{2}{c|}{ASU Campus} & 3.5 GHz & 20 MHz & $1\times 16$ & Isotropic \\
\bottomrule
\end{tabular}
\end{table}

\begin{figure}[t]
\vspace{-0.1cm}
\centering
\includegraphics[width=0.95\columnwidth]{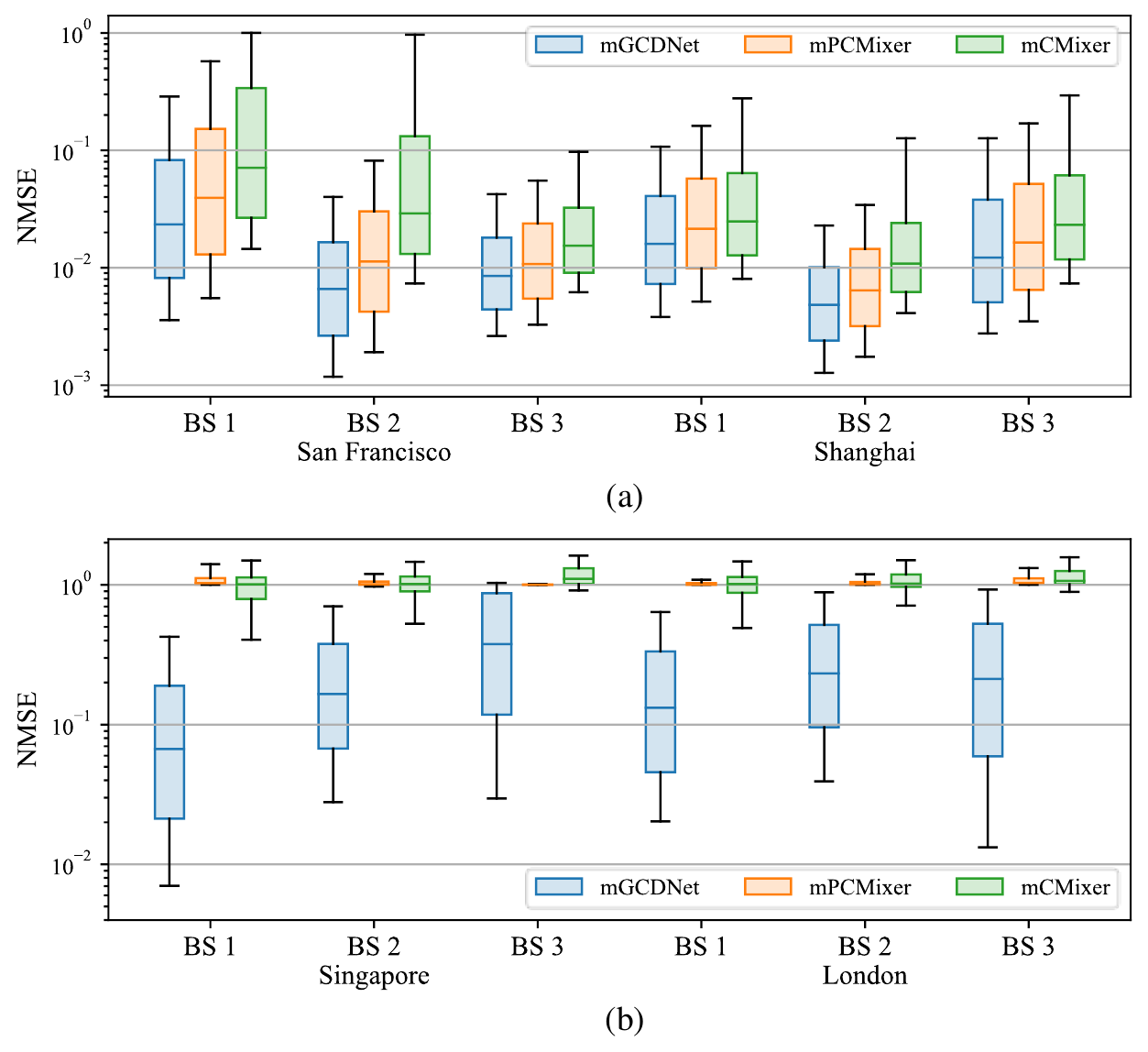}
\caption{NMSE performance in multi-scenario learning. (a) NMSE box plots in learned scenarios. (b) NMSE box plots in unseen scenarios without additional training.}
\label{fig::results_all_2}
\vspace{-0.1cm}
\end{figure}

\begin{table}[t]
\aboverulesep=0pt
\belowrulesep=0pt
\renewcommand{\arraystretch}{1.2}
\centering
\caption{NMSE and $\rho$ performance in three example scenarios.}
\label{tab::results_sfsg_asu}
\begin{tabular}{c|c|cc|cc}
\toprule
\multirow{2}{*}{Scenario} & \multirow{2}{*}{Scheme} & \multicolumn{2}{c|}{NMSE (dB)} & \multicolumn{2}{c}{$\rho$} \\
& & Mean & Median & Mean & Median \\
\midrule
\multirow{3}{*}{\makecell[c]{San Francisco\\BS 1}}
& mGCDNet & \textbf{-9.75} & \textbf{-16.30} & \textbf{0.9568} & \textbf{0.9924} \\
& mPCMixer & -7.70 & -14.05 & 0.9291 & 0.9876 \\
& mCMixer & -5.44 & -11.49 & 0.8576 & 0.9772 \\
\midrule
\multirow{3}{*}{\makecell[c]{Singapore\\BS 1}}
& mGCDNet & \textbf{-8.14} & \textbf{-11.74} & \textbf{0.9399} & \textbf{0.9815} \\
& mPCMixer & 0.50 & 0.12 & 0.1579 & 0.1030 \\
& mCMixer & -0.09 & 0.04 & 0.4783 & 0.4613 \\
\midrule
\multirow{3}{*}{ASU Campus}
& mGCDNet & \textbf{-4.85} & \textbf{-7.37} & \textbf{0.8540} & \textbf{0.9555} \\
& mPCMixer & 0.58 & 0.08 & 0.1882 & 0.1172 \\
& mCMixer & 0.75 & 0.35 & 0.2418 & 0.1336 \\
\bottomrule
\end{tabular}
\end{table}

\begin{figure}[t]
\vspace{-0.1cm}
\centering
\includegraphics[width=0.95\columnwidth]{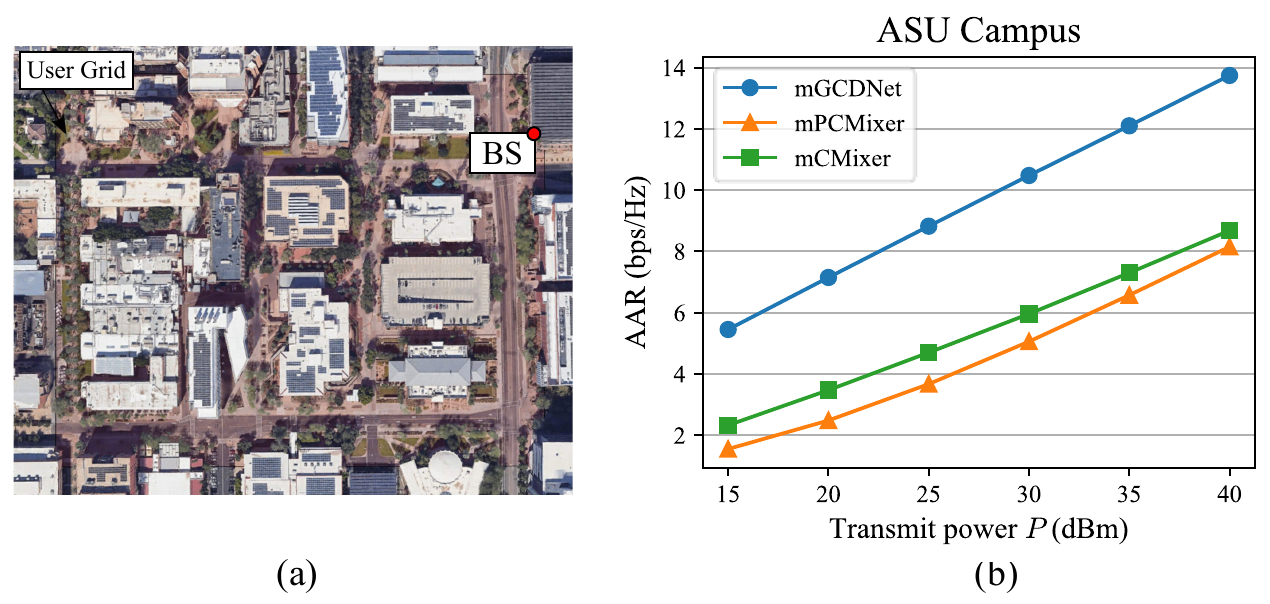}
\caption{Experiments on ASU Campus. (a) Satellite image of the scenario. (b) Mean AAR under different BS transmit power values.}
\label{fig::results_asu}
\vspace{-0.1cm}
\end{figure}

\subsubsection{Generalization across System Configurations}

To evaluate the generalization capabilities across different system configurations, we generate another group of channel datasets, where each scenario has its own carrier frequency, bandwidth, BS antenna array geometry, and antenna pattern. We first employ Sionna RT to regenerate channel data based on the 12 scenarios with new system configurations. We enable the propagation paths with no more than 5 reflections. Additionally, to involve different ray tracing platforms and settings, we introduce the ASU Campus scenario from DeepMIMO \cite{alkhateeb2019deepmimo}, an outdoor scenario with channel data simulated using Wireless Insite. The satellite image of this scenario is shown in Fig. \ref{fig::results_asu}(a). The simulation considers paths containing up to 6 reflections, 1 diffraction, or 1 scattering event. The detailed system configurations of these scenarios are shown in Table \ref{tab::dataset_x}. We train each scheme in the first 6 scenarios and evaluate directly across all 13 scenarios, including the newly added ASU Campus.

Fig. \ref{fig::results_all_2}(a) shows the NMSE performance across the 6 scenarios involved in training. mGCDNet consistently outperforms the other schemes. Fig. \ref{fig::results_all_2}(b) shows the NMSE performance across 6 unseen scenarios. The baseline schemes completely fail in these new scenarios, while mGCDNet exhibits generalization capabilities on part of the testing samples, especially in Singapore BS 1. These testing samples may be covered by the data distribution from previous training scenarios, which is why the proposed method works. Thus, by enriching the training data and increasing scenario diversity to achieve broader coverage of the data distribution, the proposed method is expected to generalize to more testing samples in new scenarios and improve the overall performance. Alternatively, computation-efficient finetuning can be performed using data with new system configurations. 

Fig. \ref{fig::results_asu}(b) shows the AAR performance of different schemes in ASU Campus. Similar to the results in Sionna RT scenarios, mGCDNet demonstrates significant advantages over baseline schemes. Table \ref{tab::results_sfsg_asu} presents the quantitative results of different schemes in three example scenarios, further validating the superiority of the proposed mGCDNet.

\begin{figure}[t]
\vspace{-0.1cm}
\centering
\includegraphics[width=\columnwidth]{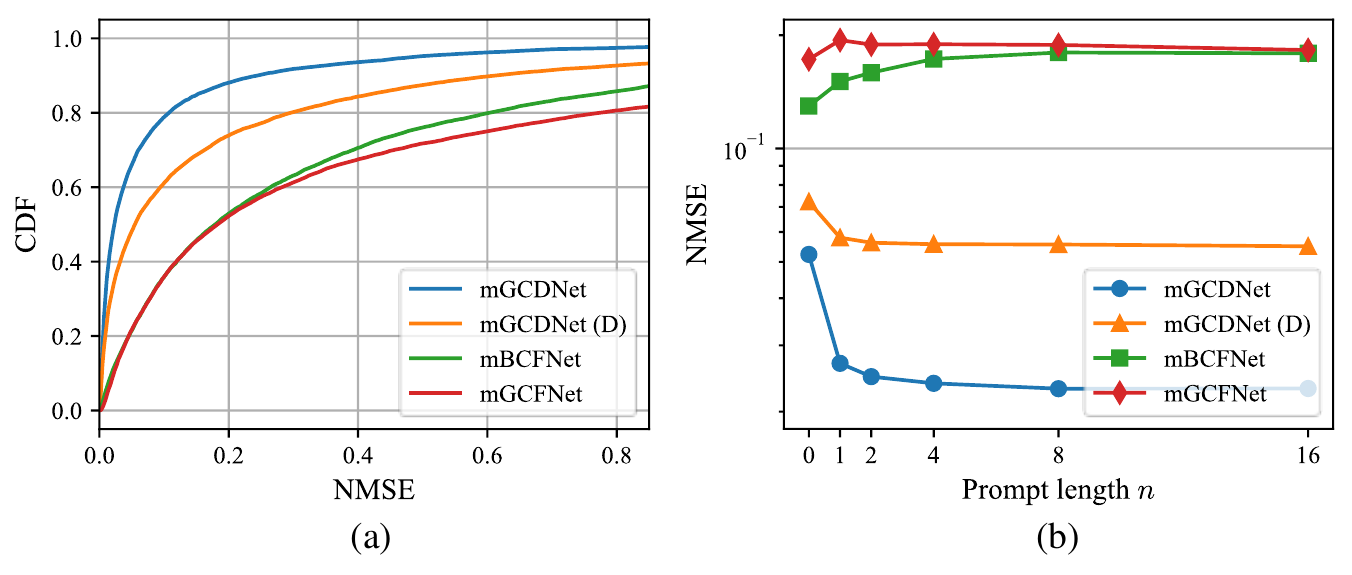}
\caption{Single-scenario learning performance of schemes with different prompt designs. (a) Cumulative probability distribution of NMSE. (b) Median NMSE under different prompt lengths.}
\label{fig::ablations_sf1}
\vspace{-0.1cm}
\end{figure}

\begin{figure}[t]
\vspace{-0.1cm}
\centering
\includegraphics[width=\columnwidth]{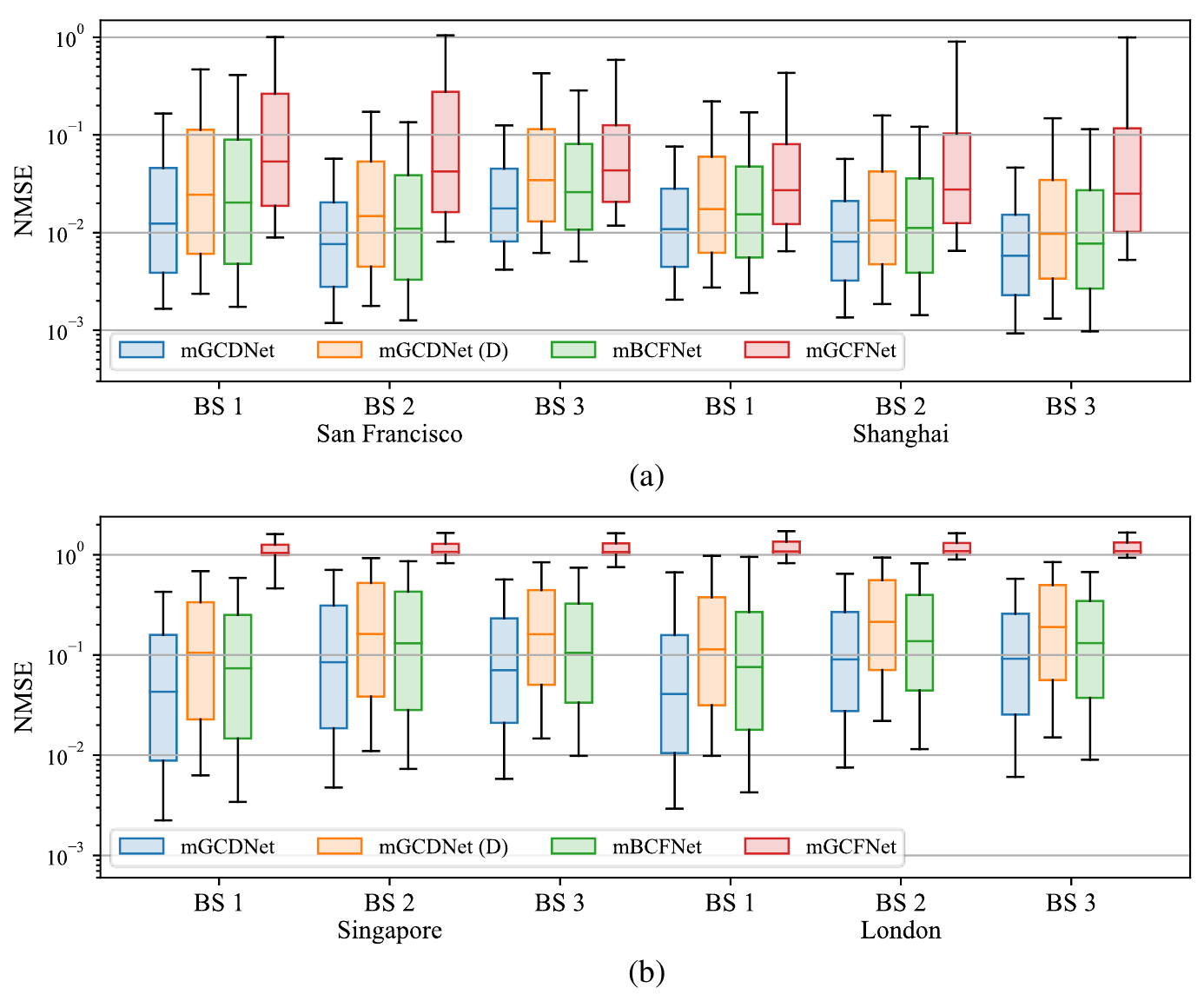}
\caption{NMSE performance of schemes with different prompt designs in multi-scenario learning. (a) NMSE box plots in learned scenarios. (b) NMSE box plots in unseen scenarios without additional training.}
\label{fig::ablations_all}
\vspace{-0.1cm}
\end{figure}

\begin{figure}[t]
\vspace{-0.1cm}
\centering
\includegraphics[width=\columnwidth]{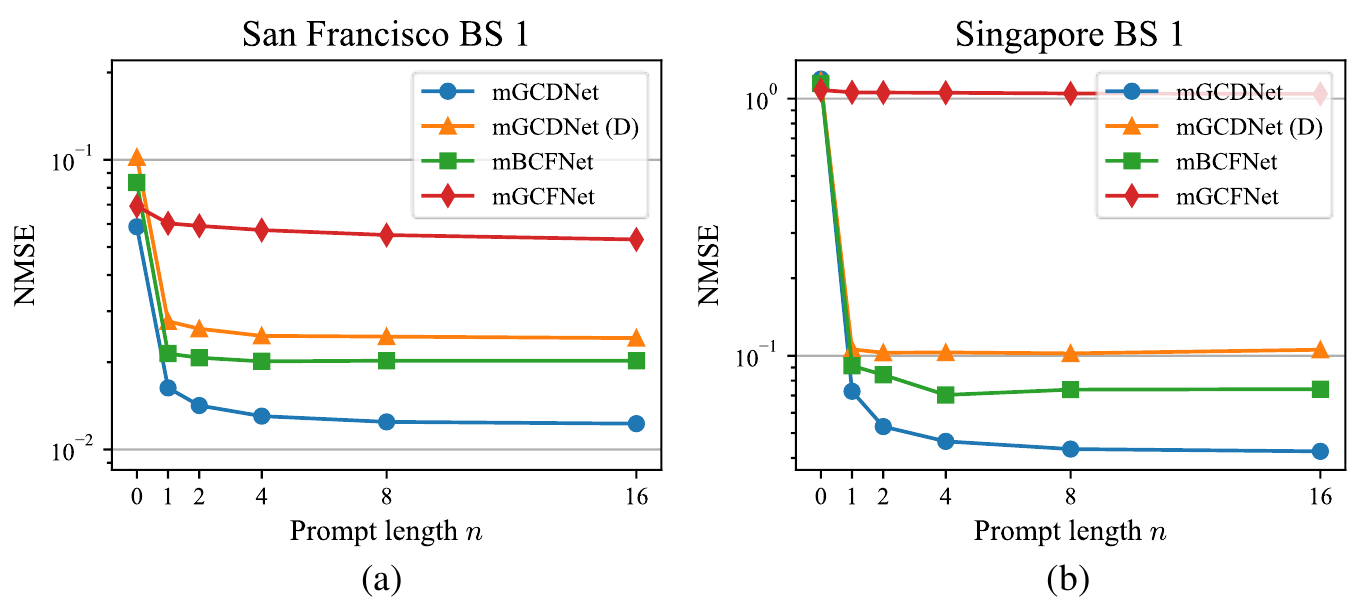}
\caption{Impact of prompt length on NMSE performance of schemes with different prompt designs. (a) Median NMSE in a learned scenario. (b) Median NMSE in an unseen scenario.}
\label{fig::ablations_sf1_sg1}
\vspace{-0.1cm}
\end{figure}

\subsection{Ablation Studies on Prompt Design}

In this subsection, we design several alternative approaches to incorporate the geometric information and compare them with the proposed random prompt augmentation method. All these methods follow the same framework -- fusing partial CSI with geometric features to acquire full CSI. The only difference among them lies in how the geometric features are handled. Through this comparison, we demonstrate that the proposed prompt design not only outperforms traditional channel estimation methods but also excels at leveraging geometric information compared with other geometry-aided methods. Specifically, we devise the following methods:

\begin{itemize}
\item \textbf{Geometry-channel fusion (GCF):} This scheme allows the neural network to learn on its own how to convert the geometric path parameters into an appropriate prompt that can be fused with the partial CSI. Since the number of paths is uncertain and there is no specific order among the different paths, we employ a Transformer without positional embedding to process the path parameters $\mathcal{P}_i$, followed by average pooling to obtain a latent feature vector $\bm{p}_i\in\mathbb{R}^D$. We use a shared Transformer to process the geometric features of all $n$ neighbors, thereby obtaining $n$ feature vectors $\bm{p}_{i_1},\cdots,\bm{p}_{i_n}$. These feature vectors serve as the contextual prompt and are fed into the channel information fusion module. We refer to this scheme as mGCFNet.
\item \textbf{Basis-channel fusion (BCF):} This scheme draws inspiration from several existing studies on DT-aided channel acquisition \cite{del2025bayesian,cai2026ecbp2wcp}. We select the path parameters $\mathcal{P}_i$ of the nearest neighbor and compute a set of matrices $\{\bm{\Phi}_{i,p}\}_{p=1}^{N_{{\rm p},i}}$, with each matrix $\bm{\Phi}_{i,p}\in\mathbb{C}^{N_{\rm t}\times N_{\rm c}}$ defined as $\bm{\Phi}_{i,p}[n_{\rm t},n_{\rm c}]=\phi_{i,n_{\rm t},n_{\rm c},p}$ (see \eqref{eq::phi_i_element_p}). These matrices form a basis (which may be non-orthogonal) of the channel subspace. We feed these $N_{{\rm p},i}$ basis matrices as the contextual prompt into the channel information fusion module. We refer to this scheme as mBCFNet.
\item \textbf{Deterministic pseudo channel construction:} This method is similar to the proposed prompt design, with the difference being that the placeholder is not random. Specifically, we construct a deterministic pseudo channel $\widetilde{\mathbf{H}}_i^{\rm(D)}\in\mathbb{C}^{N_{\rm t}\times N_{\rm c}}$ by replacing the unknown $\widetilde{\bm{\alpha}}_i$ with a placeholder $\widetilde{\bm{z}}_i^{\rm(D)}$:
\begin{equation}
\widetilde{\mathbf{H}}_i^{\rm(D)}[n_{\rm t},n_{\rm c}]=\bm{\phi}_{i,n_{\rm t},n_{\rm c}}^{\mathsf{T}}\widetilde{\bm{z}}_i^{\rm (D)},
\end{equation}
where each element of $\widetilde{\bm{z}}_i^{\rm (D)}$ is
\begin{equation}
\widetilde{z}_{i,p}^{\rm (D)}=\frac{\lambda}{4\pi d_{i,p}}z_{i,p}^{\rm (D)},\quad z_{i,p}^{\rm (D)}=\frac{\sqrt{\pi}}{2}\sigma_{\rm z}.
\end{equation}
Compared with \eqref{eq::tilde_z_element}, the variable $z_{i,p}^{\rm(D)}$ here is a constant value instead of a random one. We set this value as the expected magnitude of the original random variable, i.e., $z_{i,p}^{\rm (D)}=\mathbb{E}[|z_{i,p}|]$, with $z_{i,p}\sim\mathcal{CN}(0,\sigma_{\rm z}^2)$ as defined in \eqref{eq::tilde_z_element}. We refer to mGCDNet with this prompt design as mGCDNet (D).
\end{itemize}

All these schemes adopt the same training setting as the proposed mGCDNet. We first train each scheme in a single scenario -- San Francisco BS 1. Fig. \ref{fig::ablations_sf1}(a) shows the CDF of NMSE. We can observe that mGCDNet performs best while mGCDNet (D) ranks second. Both mBCFNet and mGCFNet perform poorly. Fig. \ref{fig::ablations_sf1}(b) investigates the impact of prompt length on performance. For mBCFNet, a prompt length of $n$ means retaining the basis matrices corresponding to at most $n$ shortest paths; for the other three schemes, the prompt length represents the number of neighboring samples. We observe that as the prompt length increases, the performance of mBCFNet and mGCFNet actually declines. This suggests that these two schemes fail to effectively utilize the auxiliary geometric information within the prompt.

Next, we train each scheme in the first 6 scenarios and evaluate across all 12 scenarios. The results are shown in Fig. \ref{fig::ablations_all}. Compared with the single-scenario case, the performance of mBCFNet and mGCFNet improves due to multi-scenario learning, and mBCFNet even outperforms mGCDNet (D). Fig. \ref{fig::ablations_sf1_sg1}(a) and Fig. \ref{fig::ablations_sf1_sg1}(b) illustrate the impact of prompt length on performance in learned and unseen scenarios, respectively.
In the learned scenario, the performance of all schemes improves as the prompt length increases, indicating that all schemes can utilize the contextual information within the prompt. This contrasts with the results in single-scenario learning in Fig. \ref{fig::ablations_sf1}(b), likely because these schemes have to learn to exploit the contextual information within the prompt during training to accommodate multiple scenarios. In the unseen scenario, mGCDNet, mGCDNet (D), and mBCFNet perform well with the assistance of contextual prompts, whereas mGCFNet completely fails. This may be because mGCFNet does not leverage the channel model and can only learn to convert the geometric information to CSI prompt based on training data, thus lacking generalization capabilities in new scenarios where the data distribution changes drastically.

Overall, mGCDNet, mGCDNet (D), and mBCFNet incorporate the channel model into prompt design, successfully achieving multi-scenario collaborative learning and cross-scenario generalization. Among these schemes, mGCDNet achieves the best performance in both single-scenario and multi-scenario cases, validating the effectiveness of the proposed random prompt augmentation method.

%% file: SecV.tex
In this paper, we propose a channel acquisition framework named GCD, which extracts geometric prior information from the DT and fuses it with pilot-based coarse estimates. We carefully design the method for scenario prompt generation and augmentation by considering the physical structure of channel data. For channel information fusion, we draw upon advanced learning architecture while enhancing it to support variable pilot configurations. Comprehensive experimental results validate the superiority of the proposed schemes, which achieve high-quality channel acquisition with low pilot overhead, strong robustness to biased geometric information, and significant cross-scenario generalization capability.

Our work highlights the importance of prompting and pre-training for wireless models. By providing necessary and accessible prompts -- such as environmental geometry, pilot patterns, and system configurations -- the model can undergo thorough pre-training using rich data from multiple scenarios with variable pilot patterns and diverse system configurations, thereby learning universal knowledge and in turn enhancing the model's intelligence. Furthermore, modality transformation and network architecture design that adhere to the physical structure of wireless channels also contribute to improving the model's learning efficiency. These insights may serve as references for future wireless model design.